\documentclass[acmsmall,screen,nonacm]{acmart}
\AtBeginDocument{%
  \providecommand\BibTeX{{%
    \normalfont B\kern-0.5em{\scshape i\kern-0.25em b}\kern-0.8em\TeX}}}
\setcopyright{acmcopyright}
\copyrightyear{2024}
\acmYear{2024}

\acmJournal{CSUR}

\newcommand{\tabincell}[2]{\begin{tabular}{@{}#1@{}}#2\end{tabular}}  

\usepackage{subfigure}
\usepackage{supertabular}
\usepackage{tabu}
\usepackage{booktabs,tabularx} %
\usepackage{multirow}

\usepackage{makecell}
\newcommand{\der}{\operatorname{d\!}{}}

\newenvironment{packed_itemize}{
\begin{list}{\labelitemi}{\leftmargin=1.5em}
  \setlength{\itemsep}{1pt}
  \setlength{\parskip}{0pt}
  \setlength{\parsep}{0pt}
  \setlength{\headsep}{0pt}
  \setlength{\topskip}{0pt}
  \setlength{\topmargin}{0pt}
  \setlength{\topsep}{0pt}
  \setlength{\partopsep}{0pt}
}{\end{list}}

\newcommand{\para}[1]{{\vspace{2pt} \bf \noindent #1 \hspace{0.5pt}}}

\begin{document}

\title{Multi-Scale Simulation of Complex Systems: A Perspective of Integrating Knowledge and Data}

\author{Huandong~Wang}
\email{wanghuanong@tsinghua.edu.cn}
\orcid{0000-0002-6382-0861}
\author{Huan~Yan}
\email{yanhuanthu@gmail.com}
\orcid{0000-0001-9626-5676}
\author{Can~Rong}
\email{rongcan@pku.edu.cn}
\orcid{0000-0002-5846-724X}
\author{Yuan~Yuan}
\email{y-yuan20@mails.tsinghua.edu.cn}
\orcid{0000-0003-1701-2588}
\author{Fenyu~Jiang}
\email{jiangfenyu@tsinghua.edu.cn}
\orcid{0000-0001-9912-9709}
\author{Zhenyu~Han}
\email{hanzy19@mails.tsinghua.edu.cn}
\orcid{0000-0001-9634-7962}
\author{Hongjie~Sui}
\email{huihj21@mails.tsinghua.edu.cn}
\orcid{0009-0007-8702-234X}
\author{Depeng~Jin}
\email{jindp@tsinghua.edu.cn}
\orcid{0000-0003-0419-5514}
\author{Yong~Li}
\email{liyong07@tsinghua.edu.cn}
\orcid{0000-0001-5617-1659}
\affiliation{%
  \institution{Department of Electronic Engineering, Beijing National Research Center for Information Science and Technology (BNRist), Tsinghua University}
  \country{China}
  }

\begin{abstract}
Complex system simulation has been playing an irreplaceable role in understanding, predicting, and controlling diverse complex systems. In the past few decades, the multi-scale simulation technique has drawn increasing attention for its remarkable ability to overcome the challenges of complex system simulation with unknown mechanisms and expensive computational costs. In this survey, we will systematically review the literature on multi-scale simulation of complex systems from the perspective of knowledge and data. Firstly, we will present background knowledge about simulating complex systems and the scales in complex systems. Then, we divide the main objectives of multi-scale modeling and simulation into five categories by considering scenarios with clear scale and scenarios with unclear scale, respectively. After summarizing the general methods for multi-scale simulation based on the clues of knowledge and data, we introduce the adopted methods to achieve different objectives. Finally, we introduce the applications of multi-scale simulation in typical matter systems and social systems.
\end{abstract}


\maketitle

\section{Introduction}

Complex systems refer to a category of systems with significant complexity, which is a concept with broad boundaries covering numerous fields of science~\cite{han2017deep,brooks2009charmm,st2021universal,xing2021multi}. For example, the human brain composed of a large scale of neurons, the physical system composed of a large scale of particles, and the transport system composed of a large scale of roads and vehicles are all representative complex systems. Their complexity often arises from the existence of the vast number of elements and components within complex systems, and nonlinearity, uncertainty, feedback loops, and emergent properties caused by the complicated interaction and co-evolution among them~\cite{karsenti2008self,battiston2021physics}. At the same time, in the modern age, numerous cutting-edge technology applications, including drug design, weather forecasting, chemical engineering, and socio-economic governance, are highly relevant to complex systems, and require effective understanding, predicting, and controlling complex systems. However, the complicated properties within complex systems often prevent us from effectively understanding, predicting, and controlling them. For example, the strong uncertainty of the current states of complex systems makes it difficult to predict their future states accurately. In addition, due to the existence of nonlinearities in the system, the aggregate effect of the interventions we impose on the system becomes no longer a simple summation of the effects of individual interventions, and often is accompanied by the effect of feedback loops and emergent properties, making the control of complex systems extremely difficult.

Fortunately, the rapid development of computing technologies has enabled us to obtain unprecedented computing power. The increasing availability of high-performance computing machines with parallel processors has made it possible to numerically compute the evolution of the massive variables in complex systems and thus effectively simulate them in this way~\cite{holland1992adaptation}. On the other hand, the rapid development of artificial intelligence (AI) technologies has made data-driven AI-powered simulation a great success, among which the most representative one is the DeepMind's achievements from protein simulation~\cite{jumper2021highly} to fluid simulation~\cite{sanchez2020learning}. Overall, computer simulation has played an irreplaceable role in  understanding, predicting, and controlling complex systems by providing excellent insight and foresight, which includes but is not limited to the following:
\begin{packed_itemize}
\item \textbf{Establish key connections with first-principle simulation:}  
In many application scenarios of complex systems, we want to establish the connections between variables, structures, and properties in the systems, which is vague in most cases. Under the circumstances, we can simulate the structure and properties from the variables using first-principles simulation, i.e., ab initio simulation, to establish the connection between them. For example, given the amino acid sequence of a protein, we want to predict its function. In this case, first-principles simulation of its structure of protein folding can help us predict its functional properties to a large extent~\cite{brini2020protein}. Or when we want to understand the phase transition process of interdependent complex networks under random attacks, we can also simulate the network connectivity in the process of cascading failure, based on which we can build connections between whether a cascading collapse occurs and the probability of attack~\cite{buldyrev2010catastrophic,bashan2013extreme}.
\item \textbf{Derive future system states with predictive simulation:} 
Simulation technology, on the other hand, is mostly used to help us derive the future state of the system from the current state to provide foresight of the system, whether through kinetic equations~\cite{barzel2013universality, gao2016universal} or other artificial intelligence models such as RNN~\cite{han2021artificial, ghavamian2019accelerating, guo2021electromagnetic} or GCN~\cite{murphy2021deep,lino2021simulating}. For example, by numerical simulation of weather, we can avoid the damage of extreme weather to agriculture. Another example is that by utilizing direct numerical simulation (DNS) to solve Navier–Stokes equations, we can predict fluid dynamics~\cite{kochkov2021machine}, which is helpful in the design of airplanes, vehicles, and engines~\cite{male2019large}.
\item \textbf{Eliminate uncertainty with sampling simulation:} 
When dealing with a complex system with strong uncertainty, the high dimensionality and strong mutual dependency of random variables within the complex system make it hard to analytically understand the system. Take a system of atomic spins described by the three-dimensional Ising model~\cite{bohm2022noise} as an example. For the large number of random variables describing the spin directions of atoms in the system, even the simple task of obtaining the probability distribution of the number of atoms with different spin orientations is difficult to be analytically solved. Under this circumstance, sampling simulation, e.g., Monte Carlo Markov simulation~\cite{yang2019high}, can be utilized to generate samples of random variables without requiring analytical solutions to their probability distributions. In this way, we can characterize the system in terms of statistics of these samples, e.g., their mean and variance, thereby eliminating the effects of the strong randomness.
\end{packed_itemize}

Despite the excellent insight and foresight provided by these methods, how to effectively simulate complex systems is still an open problem with significant challenges, which can be summarized as follows:
\begin{packed_itemize}
\item \textbf{Unknown mechanisms:} In many cases, the mechanisms of the complex systems that we want to simulate are unclear, incomplete, or difficult to obtain. A direct way to solve this problem is to utilize data-driven methods to fit the unknown mechanisms based on observable data. However, this solution has two limitations. First, the data-driven method struggles with generalization. That is, the performance of the fitted dynamics cannot be guaranteed in the scenario not covered by the observable data~\cite{kochkov2021machine}. On the other hand, the observable data are also often incomplete and noisy, which makes fitting the mechanisms very difficult.
\item \textbf{Expensive computational cost:} Even though the dynamics of complex systems are clear and complete, calculating the evolution of extremely high-dimensional state variables in complex systems is usually costly. For example, an accurate direct-numerical simulation of fluid dynamics based on Navier–Stokes equations is impossible at the scale required for solving weather prediction or engine design~\cite{kochkov2021machine}. How to simulate complex systems with acceptable computational costs is another challenge we have to face.
\end{packed_itemize}

The technique of multi-scale simulation, which has been investigated for several decades, can help us solve these two challenges. Specifically, the core idea of multi-scale simulation is to utilize the mechanisms corresponding to multiple scales simultaneously to effectively simulate the dynamics of complex systems. The different scales usually have different spatial resolutions, temporal resolutions, or quantities, and thus sometimes have completely different mechanisms~\cite{brooks2009charmm}. As for the problem of unknown mechanisms, it is trivial that we can complement or derive the macroscale mechanisms based on the simulated microscale dynamics~\cite{weinan2011principles}. Moreover, the microscale dynamics can be better modeled with the boundary conditions or background fields~\cite{wu2020recurrent} obtained based on the simulated macroscale dynamics. Sometimes, the macroscale and microscale mechanisms are both incomplete and tightly coupled to each other, and thus we have to simulate their dynamics together in a concurrent manner~\cite{liu2019exploring}. Overall, through the method of multi-scale simulation, we can complete the incomplete mechanisms of a specific scale, and learn or calibrate the parameters of the target scale mechanism, thus overcoming the challenges of unknown mechanisms. As for the expensive computational cost, the dynamics simulated based on macroscale mechanism are more coarse-grained and less accurate, but it is significantly less computationally expensive, since the number of variables that should be handled is greatly reduced. A direct way is to use the macroscale dynamics to approximate the microscale dynamics~\cite{vlachas2022multiscale}. Furthermore, the unimportant areas can be simulated and approximated based on macroscale mechanisms, while microscale mechanisms are still used for simulating important areas~\cite{levitt1975computer}. Through appropriate message-passing mechanisms between dynamics of different scales, we can simulate the complex system with both computational efficiency and simulation accuracy, thus overcoming the challenges of expensive computational costs.

In this survey, we will systematically review the recent literature on multi-scale simulation of complex systems, which include both matter systems and social systems. Our goal is to conduct an interdisciplinary literature survey to summarize the common paradigms and methodologies of existing work on multi-scale simulation from the perspective of knowledge and data, thereby helping the computer simulation community, machine learning community, transportation society, and others. We hope this survey will help multidisciplinary communities to understand the progress made in existing work and draw useful insights and perspectives from work in other disciplines, as well as potential gaps and opportunities. To provide a comprehensive understanding of interdisciplinary work on multi-scale simulation of complex systems, we provide a novel multi-perspective taxonomy of existing work in terms of their objectives, general methods, objective-oriented methods, and applications. Furthermore, the taxonomy of methods in our paper is closely organized around the topic of how existing work leverages knowledge and data. To the best of our knowledge, no existing survey has covered existing studies in terms of both matter complex systems and social complex systems. We hope that this paper will help advance progress in the direction of multi-scale simulation of complex systems.

The structure of this paper is as follows. We begin by introducing related reviews of our paper by comparing their main difference. Then, we introduce background about the simulation of complex systems and scales in complex systems. Next, we summarize the main objectives of multi-scale modeling and simulation techniques. Following the introduction of general methods of multi-scale simulation techniques based on the clues of knowledge and data introduced in Section~\ref{sec:genmethod}, we introduce the mainly adopted methods with different objectives in Section~\ref{sec:objmethod}. Finally, we introduce the application of multi-scale simulation techniques in different systems, including major matter systems and social systems.

\section{Related Reviews}\label{sec:relatedreviews}

\begin{table*}[ht]
  \begin{center}
  \begin{tabular}{|c|p{12cm}|}
  \hline
   \textbf{Paper} & \textbf{Description} \\\hline
    \cite{gomes2018co}       &    \tabincell{l}{This paper presented the related techniques, tools, and challenges for co-simulation, \\where a coupled system can be simulated by integrating with multiple simulators.}  \\ \hline
    \cite{noe2017collective}     &   \tabincell{l}{This paper summarized the recent developments of finding the collective variables \\of a system, including the related principles, and the state-of-the-art algorithms.}   \\ \hline
    \cite{klus2018data}          &    \tabincell{l}{This paper studied different data-driven methods to project dynamics of the \\high-dimensional system into the lower-dimensional space, and further explored the \\similarity and differences between these methods. }  \\ \hline
    \cite{peng2021multiscale}         &  \tabincell{l}{This paper summarized the recent multi-scale methods that integrate with machine \\learning, and discussed the mutual benefits with each other.}\\ \hline
    \cite{albi2019vehicular}      & \tabincell{l}{This paper surveyed the multi-scale methods in the cases of vehicles and crowds, \\which mainly focused on the bottom-up derivation of models.}\\ \hline
  \end{tabular}
  \end{center}
    \caption{Related reviews of complex system modeling and simulation.}
  \label{Tab:existingsurvey}
\end{table*}

There are some comprehensive articles that summarize different techniques, tools, and algorithms in the area of simulation~\cite{noe2017collective,klus2018data,gomes2018co,albi2019vehicular,peng2021multiscale}. Gong et al.~\cite{gomes2018co} gave a comprehensive review about co-simulation, where a coupled system can be simulated by the composition of different simulators. The corresponding theory and techniques were surveyed, which are grouped into two types of paradigms: discrete event co-simulation and continuous time co-simulation. Further, this work studied the hybrid co-simulation that establishes the relationship between these two paradigms. Frank et al.~\cite{noe2017collective} focused on the role of collective variables in studying high-dimension dynamical systems. The collective variables can discover meaningful collective phenomena over a long period of time, as well as identify different structures at macroscopic scales. In the system simulation, it is important to find the collective variables of the system to control the simulation. The authors reviewed the relevant theory and algorithms that search for the collective variables from simulation data. Similarly, Klus et al.~\cite{klus2018data} studied the recent data-driven methods in discovering the collective variables of a system, including data-driven dimension reduction methods and transfer operator approximation methods. Further, they summarized the similarities and differences between these methods, and discussed how extensions designed for a specified method can be applied to other related ones. Peng et al.~\cite{peng2021multiscale} presented the multi-scale modeling methods based on physics-based knowledge and data-driven machine learning. They discussed how multi-scaling modeling integrates with machine learning, and explained the mutual benefits with each other. Albi et al.~\cite{albi2019vehicular} surveyed the multi-scale methods in the cases of vehicles and crowds, where the models at the macroscopic scale are derived from individual-based modeling at the microscopic scale. Different from them, our paper gives a comprehensive analysis of multi-scale simulation from different aspects, but the above articles point out a small part of them. To be specific, we identify three interaction ways in the scenarios of clear-scale modeling based on the flow of information between scales, including information flow from microscale to macroscale, from macroscale to microscale, and the mutual interaction between microscale and macroscale. However, \cite{albi2019vehicular} only discussed the interaction from microscale to macroscale, and \cite{noe2017collective} and~\cite{klus2018data} focused on the interaction from macroscale to microscale. Furthermore, our paper considers the scenarios of multi-scale simulation with unclear scales in complex systems, and two additional objectives of multi-scale simulation, including {\em discovering unknown scales} and {\em learning dynamics at the discovered scale}. Therefore, this paper can provide more insightful and valuable information for researchers to understand the multi-scale simulation methods in complex systems.

\section{Background}

To pave the way for reviewing the literature on multi-scale simulation of complex systems, we first introduce background about the concepts of simulation and scales of complex systems, respectively.

\subsection{Simulation of complex systems}\label{sec:background_scales}

\begin{figure}[t]
	\centering
    \includegraphics[width=0.5\textwidth]{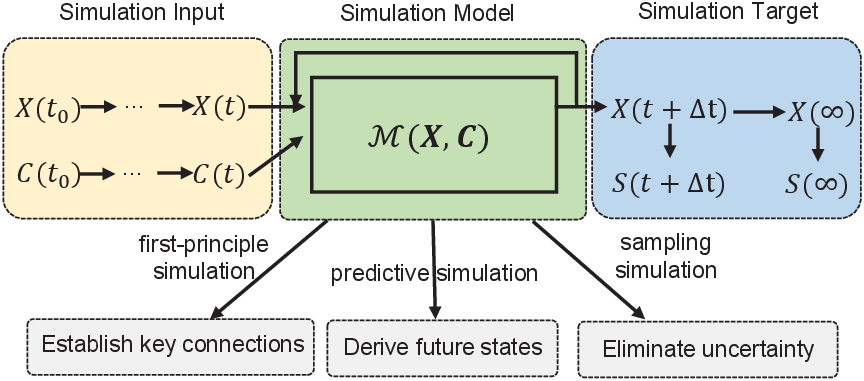}
	\caption{An illustration of simulation of complex systems, where $X(t)$ is the observable system state variables at time $t$, $C(t)$ is the controllable condition variable of the system. $\boldsymbol{X}$ and $\boldsymbol{C}$ are the set of system state variables and controllable condition variables from $t_0$ to $t$, respectively. $S(t)$ is the target variable that we want to obtain through simulation.}
\label{fig:simulate}
\end{figure}

The goal of the simulation of complex systems is to imitate the operation or behavior of the systems. As shown in Figure~\ref{fig:simulate}, by denoting the observable system state variables at time $t$ as $X(t)$ and the controllable condition variables as $C(t)$, a simulation model $\mathcal{M}(\boldsymbol{X},\boldsymbol{C})$ is built to mimics the target complex system. Specifically, we define $\boldsymbol{X}=\{X(\tau)\}^t_{\tau=t_0}$ and  $\boldsymbol{C}=\{C(\tau)\}^t_{\tau=t_0}$, which are the set of all historical observable system state variables and controllable condition variables, respectively.

In some systems, the simulation models only take $X(t)$ and $C(t)$ as inputs. It depends on whether the system is Markovian, i.e., whether the future state depends only on the currently observed system state variables. This is true for a large number of physical systems. For example, in a fluid system of gases, given the current velocity field and external forces, the future state, i.e., the future velocity field, can be accurately simulated. However, this is not true for most social systems, such as transportation systems. The essential reason is that the observable state variables are not enough to completely determine the state of the system, and there exist unobservable hidden state variables in the system. For example, in a transportation system, we can observe the position and speed of each vehicle, but this is not enough to fully determine the future positions and speeds of all vehicles. In addition to the mental states of the vehicle drivers, each vehicle also has a different destination, which is hidden in the system and largely affects the future vehicle behavior. Under these circumstances, the historical sequence of observable state variables should be introduced in the simulation model to estimate the hidden state of the systems.

The simulation target is diverse, and depends on which aspects of the simulation model are required to imitate the operation or behavior of the real complex system. For example, in different simulations, different target variables $S(t)$ are required to be simulated. In addition, sometimes the whole simulated process of system state evolution, i.e., $\{S(t+\Delta t)|\Delta t\in(0,\infty)\}$ is required to imitate the actual complex system, while other times we only focus on the final state of the system $S(\infty)$. At the same time, sometimes we need to simulate the exact value of the system state, i.e., $S=h(X)$, and other times we need to estimate its probability distribution $S=Pr(h(X))$.

\subsection{Scales of complex systems}

\begin{figure}[t]
	\centering
	\subfigure[Typical clear scales in complex systems]{\label{fig:rules_tran}\includegraphics[width=.4\textwidth]{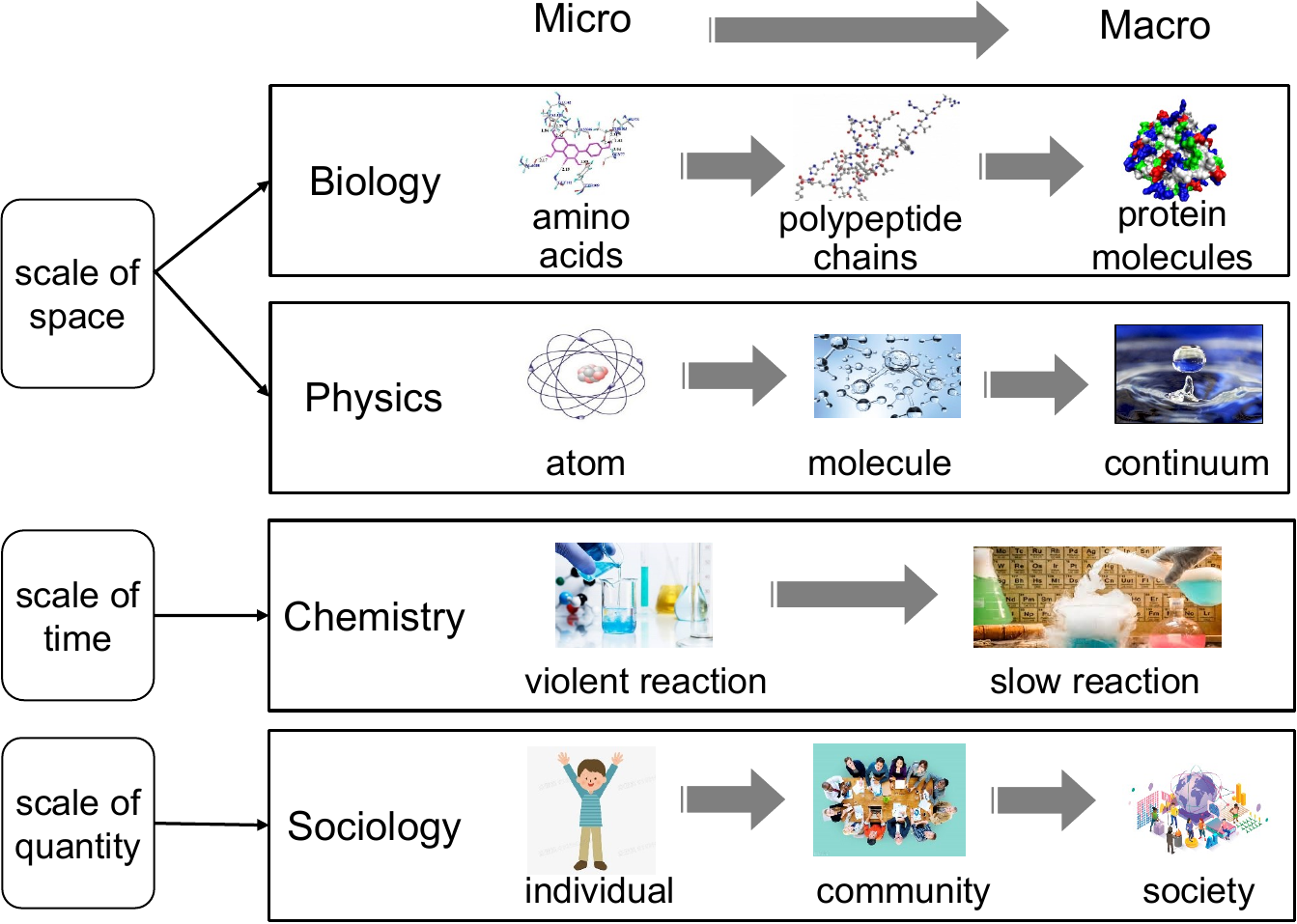}}
	\hspace{0.2cm}
   \subfigure[Numerous unclear scales in complex systems]{\label{fig:rules_visit} \includegraphics[width=.5\textwidth]{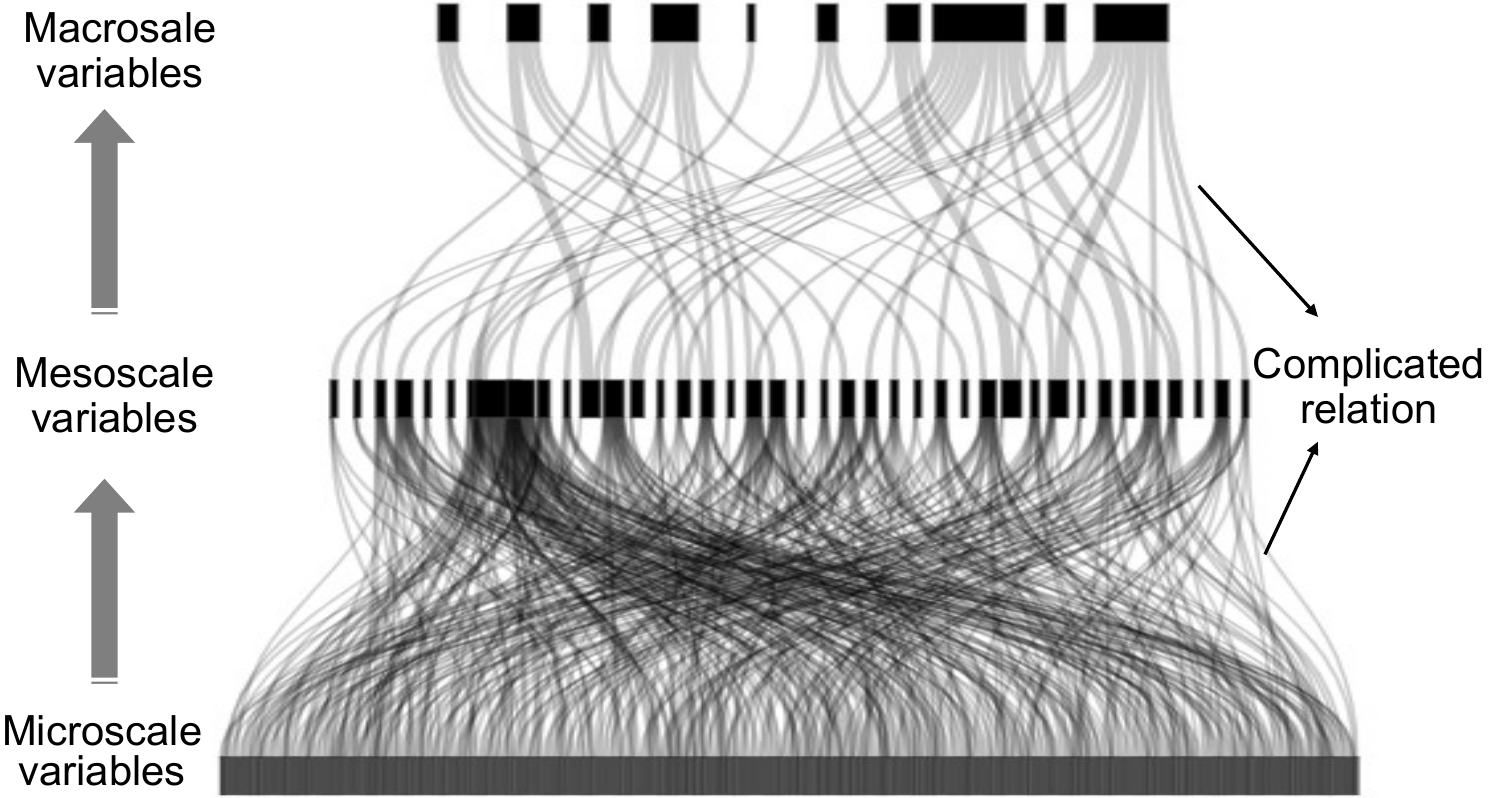}}
	\caption{An illustration of scales in complex systems.}
\label{fig:scales}
\end{figure}

There exist various categories of scales in complex systems. As shown in Figure~\ref{fig:scales}(a), typical scales can be summarized as the following three categories. The first category is the scale of space, of which the definition is diverse in different scientific fields. For example, in physics, from electrons to atoms, further to molecules, and continuum, the corresponding spatial scales range from a few nanometers to a few meters. In biology, it starts from amino acids to polypeptide chains, and finally forms protein molecules. The second category is the scale of time. A representative example is that there exist violent reactions and slow reactions in chemical reaction complex systems, which lead to the formation of fast and slow variables in the system. The last category is the scale of quantity. For example, in society, there exist scales from the individual to the community, and finally to the society. Note that scales can generally be divided into two types. In the first type of scales, the corresponding state is defined over a continuous area, such as the velocity field of a fluid. In the second type of scales, the corresponding state is defined based on a limited number of variables, such as vehicles in a transportation system, and nodes in a complex network. We refer to the former type of scales as the field-based scale and the latter as the particle-based scale.

Despite the typical scales with clear definitions and physical meanings in complex systems, there also exist numerous unclear scales in complex systems. As shown in Figure~\ref{fig:scales}(b), the massive microscale variables in the system form mesoscale variables through complicated combinations and transformations, and further form macroscale variables in the system. These scales often do not correspond to a uniform space size, temporal length, or quantity, but correspond to the emergent structures~\cite{hoel2013quantifying,li1994particle} or laws~\cite{west1999fourth} in the complex system, and play an irreplaceable role in the accurate simulation of the future evolution of the system. In most cases, these scales cannot be effectively extracted based on simple space size, time length, or quantity scale. Under the circumstances, techniques such as manifold learning, and representative learning are required to be utilized to find the hidden scales in complex systems.

\section{Objectives of multi-scale modeling and simulation}

We first summarize the main objectives of existing techniques of multi-scale simulation. As discussed in Section~\ref{sec:background_scales}, there exist clear scales corresponding to different space sizes, temporal lengths, and quantities in complex systems, and numerous unclear scales as well. The clarity of the system's useable scales also results in differences in the objectives of the corresponding multi-scale simulation approaches. Thus, in the following, we will introduce the objectives of multi-scale simulation in the scenario with clear scales and unclear scales, respectively. Without loss of generality, we denote the microscale system state variables as $X(t)$, and denote the macroscale system state variables as $\xi(t)$ in this section.

\subsection{Scenario with clear scale}

Although the clear-scale modeling and simulation of complex systems are commonly adopted in many fields, such as physics~\cite{lino2022remus,lino2022towards,fortunato2022multiscale,lino2021simulating,karniadakis2021physics,xie2018crystal,faraji2017high}, material science~\cite{liang2018deep,le2015computational,liu2019deep,wang2019coarse,fish2021mesoscopic}, chemistry~\cite{han2017deep,brooks2009charmm}, and bioprotein~\cite{bhatia2021machine,ingolfsson2022machine,hambli2011multiscale}. The scale separation can vary in different fields. In the field of fluid/solid state physics, it is often necessary to focus on atomic and above scales modeled using classical mechanics and microscopic electronic scales modeled using quantum mechanics~\cite{brooks2009charmm}. In materials science, researchers usually focus on two scales, i.e., continuum strains and atomic movements, due to the challenges posed by the heterogeneity of materials~\cite{fish2021mesoscopic}. In the study of biological proteins, different scale separations are done in different research scenarios, such as atomic level in the discovery of new protein structures, and coarse-grained lipids in the study of membrane remodeling~\cite{bhatia2021machine,ingolfsson2022machine}. All the above complex system simulations are explicitly divided into multiple scales, and each scale holds a specific spatio-temporal resolution. Consistent with the existing works, in the clear scale scenario of the existing literature, the scales are ordered according to the coarseness of the spatio-temporal granularity.

Specifically, we divide the main objectives of multi-scale simulation into three categories, which correspond to different ways of interaction between different scales. It is worth noting that the methods do not need to be restricted to sequential modeling or concurrent modeling~\cite{weinan2011principles}. What matters is whether the flow of information between scales is from macroscale to microscale or vice versa. The introduction of three objectives is as follows.
\begin{packed_itemize}
    \item \textbf{Resolving local fine-grained dynamics (\textit{Objective A}):} In this objective, the local fine-grained microscale dynamics are what the study is interested in. That is, the simulation target depends only on a local subset of the system state variables, i.e., $S(t)=h(X_{l}(t))$ where  $X_{l}(t)\subset X(t)$.
    The microscale models are leveraged to resolve the local dynamics $X_{l}(t)$ in detail, while the macroscale models are utilized to simulate the coarse-grained patterns anywhere else, i.e.,  $X(t)/X_{l}(t)$. 
    \item \textbf{Completing macroscale mechanisms (\textit{Objective B}):} \textit{Objective B} is the exactly opposite-type of \textit{Objective A}. The macroscale dynamics of systems are what we care about. That is, the simulation target $S(t)=h(\xi(t))$. However, the macroscale mechanism is incomplete. For example, some constitutive information is missing, which is then obtained from microscale models and serves as the key components of macroscale models. 
    \item \textbf{Co-simulating microscale and macroscale dynamics (\textit{Objective C})}. To achieve this objective, there is a mutual interaction between the microscale models and the macroscale models. The dynamics of a specific scale or its derived statistics will be updated through the models of the corresponding scale and then passed to the other. The simulation of the complex system will be a process of iterative computation of the macroscale models and the microscale models and the information interaction between them. 
\end{packed_itemize}
\subsection{Scenario with unclear scale}

\begin{figure}[t]
	\centering
    \includegraphics[width=0.4\textwidth]{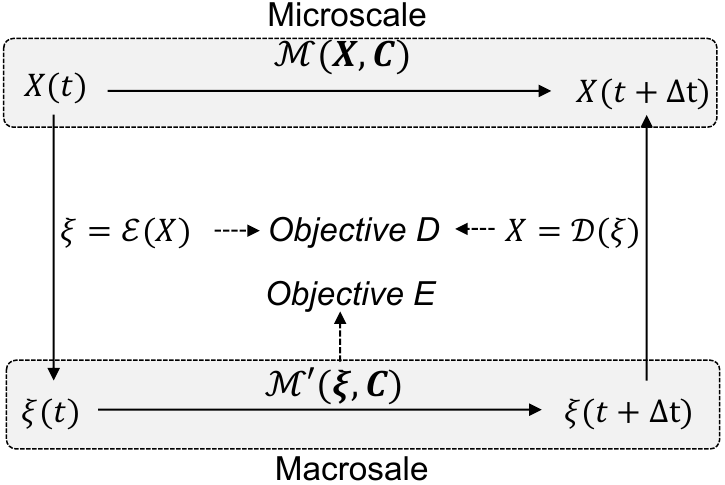}
	\caption{An illustration of {\em objective D} and {\em objective E}.}
\label{fig:objDE}
\end{figure}

Although classic fields, such as physics, chemistry, and biology, have proposed effective clear-scale models, the useable scales in complex systems are not always clear in complex systems, especially for high-dimensional systems.
Here, a fundamental problem is which scales are more appropriate for conducting multi-scale simulations for a particular target system.
For example, in a large range of chemical systems, where different reactions compete with each other, it is difficult to study why reactions occur and how to predict and control their behaviors at a clear scale~\cite{crim2008chemical,rossler1976chemical}. In ecosystems, species reside and interact with each other, forming complex scales in communities. Learning the dynamics of communities to reveal how ecosystem functions are essential for ecology~\cite{hu2022emergent}. 
Complex systems cannot always be explicitly divided into clear and appropriate scales for different simulation tasks. To deal with these systems, it is necessary to automatically discover appropriate unknown scales in complex systems by reducing the original high-dimensional microscale system state to a low-dimensional macroscale system state, and learn the dynamics at the discovered scale simultaneously.
Thus, the objectives of the corresponding multi-scale simulation approaches can be further classified into two categories, of which an illustration is shown in Figure~\ref{fig:objDE} and the  corresponding explanations are as follows:

\begin{packed_itemize}
    \item \textbf{Discovering unknown scales (\textit{Objective D}):} Discovering unknown but useful scales is the key to effectively modeling complex systems without clear scales.
    This objective aims to extract more tractable scales from complex systems. Further, we must ensure the tractability and usefulness of the extracted scales by adding constraints in terms of specific properties, e.g., linearity or preserved relevant information. More specifically, as shown in Figure~\ref{fig:objDE}, this objective requires obtaining an encoder $\mathcal{E}$, which maps the microscale state variable $X$ to the discovered macroscale state variable $\xi$, and a corresponding decoder $\mathcal{D}$ to solve the inverse problem.
    \item \textbf{Learning dynamics at the discovered scale (\textit{Objective E}):} This objective is accompanied by {\em Objective D}, which aims to learn the effective dynamics $\mathcal{M}'(\boldsymbol{\xi},\boldsymbol{C})$ at the discovered target scale. More specifically, it is a variant of {\em Objective B} in the scenario with unclear scale. Unlike {\em Objective B}, in this objective, the macroscale mechanism may be completely unknown rather than incomplete. At the same time, constraints can be added to the associated problem of {\em objective E} to ensure that the macroscale mechanism to be learned has specific properties, e.g., linearity.
\end{packed_itemize}

By combining {\em Objective D} and {\em Objective E}, the high-dimensional state variables of complex systems are able to be reduced to a low-dimensional macroscale state variable system with learned dynamics. Thus, the system is able to be simulated at the discovered low-dimensional scale.

\section{General methods of multi-scale modeling and simulation}\label{sec:genmethod}

Before introducing the multi-scale simulation methods used in achieving each objective in detail, we first summarize the general methods of multi-scale simulation techniques based on whether they are knowledge-driven, data-driven, or data and knowledge jointly driven.

\subsection{Pure knowledge-driven method}
               
Early approaches mainly adopt knowledge-driven methods to model and simulate complex systems. The commonly used methods mainly include kinetic equation models and statistical probability models.

\textbf{Kinetic equation model.} This model concentrates on modeling the evolution of the system variables in the continuous time domain. Specifically, it uses the differential equation of system state variables with respect to time to model and simulate the system, which is able to display the evolution of the complex system over time intuitively. By denoting the system state variable as $X$, which is usually high-dimensional, a kinetic equation model describing the evolution of the system with respect to time $t$ can be represented by:
\begin{equation}
\frac{dX}{dt} = f(X).
\end{equation}
For example, the dynamics of a complex network are modeled by~\cite{barzel2013universality, gao2016universal}:
\begin{equation}
\frac{dx_i}{dt} = F(x_i)+\sum_{(i,j)\in E}G(x_i.x_j),
\end{equation}
where $x_i$ is the state variable for the node $i$ in the complex network, $E$ is the set of edges of the complex network, and $F$ and $G$ are two functions describing the evolution mechanism of the complex network.

Another example is the fluid dynamics, which can be modeled by the Navier–Stokes equations:
\begin{equation}\label{equ:NS-1}
    \frac{d\boldsymbol{x}}{dt} = - \bigtriangledown\cdot (x\otimes x) + \frac{1}{Re}\bigtriangledown^2x-\frac{1}{\rho}\bigtriangledown \boldsymbol{p} + f,
\end{equation}
\begin{equation}\label{equ:NS-2}
    \bigtriangledown\cdot \boldsymbol{x}=0,
\end{equation}
where $\boldsymbol{x}$ is the flow velocity. Normally, $\boldsymbol{x}$ is a field defined on three-dimensional space. $\boldsymbol{p}$ is the field of pressure.

Kinetic equations can also be utilized to describe complex systems with stochastic dynamics. In this case, stochastic variables are introduced into the system, and derive stochastic differential equations (SDE) as follows:
\begin{equation}\label{equ:SDE}
{dX} = f(X){dt} + \sigma(X)dW,
\end{equation}
where $W$ usually represents a Brownian motion, and thus its derivative represents a Gaussian noise. Specifically, SDE is utilized to model chemical reaction systems~\cite{singer2009detecting}, biological cell systems~\cite{zhou2021dissecting},  and financial systems~\cite{mcleish1997fitting}, etc.

\textbf{Probability statistical model.} In some complex systems, the state of the system is not deterministic, but a probabilistic distribution defined on the state space. Under the circumstances, the system states as well as the system dynamics are modeled based on probabilistic models, in which the most widely used model is the Markov chain (MC) model. Specifically, in quantum mechanics, the system state usually cannot be measured accurately, and thus it is often modeled by a probability density distribution. Furthermore, the SDE in the form of (\ref{equ:SDE}) also falls into the category of Markov models. More generally, the continuous-time Markov chain (CTMC) models the system in the continuous time domain, and it describes the system dynamics through the transition density:
\begin{equation}
\mathcal{T}(x, y; \tau) := p(X(t+\tau) = y | X(t) = x),
\end{equation}
where we can observe that the transition density is independent with the time $t$, and $X(t+\tau)$ and $X(t)$ can be either defined on the continuous space or the discrete space.

On the other hand, the discrete-time Markov chain (DTMC) models the system in the discretized time domain. In this case, the future state $X_t$ of the system at the time slot $t$ is only dependent on the previous state $X_{t-1}$, and independent with the system state at earlier time slots, which can be represented by:
\begin{equation}
Pr(X_t|\{X_\tau\}_{\tau< t})=Pr(X_t|X_{t-1}) =: T(X_{t-1},X_t)
\end{equation}
Representative DTMC models include the independent cascade model~\cite{saito2008prediction}, GLobal Epidemic and Mobility (GLEAM) model~\cite{balcan2009multiscale}, etc.

Similar to the kinetic equation model, the Markov chain model also focuses on modeling system evolution over time. However, in some specific applications and scenarios, we do not care about the evolution of complex systems over time. Instead, we focus on the final state of the system $X(\infty)$ or its statistics $S(\infty)$. For example, Hoel et al.~\cite{hoel2013quantifying} only consider one-step state transition of complex systems, where the considered states are referred to as the initial states and the resulting states, respectively. Further, the energy-based model (EBM), e.g., Ising model~\cite{nauenberg1974renormalization}, can be utilized to describe the relation between the system state $S$ and condition variables $C$. Specifically, by denoting the energy as $E(S,C)$, the probability of $S$ can be represented by $Pr(S)=e^{E(S,C)}/\sum_{S'\in\mathcal{S}}e^{E(S',C)}$, where $\mathcal{S}$ is the set of all possible system state.

\begin{table*}[ht]
  \begin{center}
  \resizebox{1\textwidth}{!}{
  \begin{tabular}{c|l|l|l|l}
  \hline
   \textbf{Categories} & \textbf{Method} & \textbf{Papers} &  \textbf{Considering time}  &\textbf{Deterministic/stochastic}   \\
  \hline\hline
    \multirow{2}{*}{Kinetic equation model}
    & ODE  & \cite{barzel2013universality, gao2016universal} & Yes  & Deterministic  \\ \cline{2-5}
    & SDE  & \cite{singer2009detecting}  & Yes  & Stochastic  \\ \cline{2-5}
    \cline{1-5}
  \multirow{3}{*}{Statistical probability model}
    & CTMM  & \cite{noe2013variational}  & Yes  &  Stochastic\\ \cline{2-5}
    & DTMM  & \cite{saito2008prediction,balcan2009multiscale,hoel2013quantifying}  & Yes \& No   & Stochastic   \\ \cline{2-5}
    & EBM & \cite{mehta2014exact,nauenberg1974renormalization} & No & Stochastic \\
  \cline{1-5} 
  \cline{1-5}
  \end{tabular}
  }
  \end{center}
  \caption{Knowledge-based multi-scale modeling and simulating methods}\label{Tab:architecture}
\end{table*}

\subsection{Pure data-driven method}

\subsubsection{Data-driven modeling and simulation approaches}
Since the recently rapid development of machine learning technologies, and the availability of massive experimental datasets, data-driven methodologies appear as a surrogate to the conventional numerical methods governed by the physical equations in the field of simulation. Due to the computational efficiency and the acceptable simulation accuracy, the data-driven methods are leveraged in biology, chemistry, material science, and fluid dynamics, with the purpose of single-scale or multi-scale modeling. In this section, we prefer to provide an overview of the various machine learning techniques incorporated into the simulation procedure. These approaches mainly include methods such as the Recurrent Neural Network (RNN), the Convolutional Neural Network (CNN), the Graph Neural Network (GNN), ODENet, and the Gaussian Process (GP), etc. Here we briefly introduce each method and discuss how they benefit the simulation work.

\begin{table*}[ht]
  \begin{center}
  \begin{tabular}{c|l|l|l}
  \hline
  \textbf{Categories} & \textbf{Scale} & \textbf{Method} & \textbf{Papers}\\
  \hline\hline
  \multirow{3}{*}{Single-scale simulation}
    & \multirow{2}{*}{Micro}
    & RNN & \cite{han2021artificial, ghavamian2019accelerating, guo2021electromagnetic}\\ \cline{3-4}
    & & CNN & \cite{guo2021electromagnetic}\\ \cline{2-4}
    & Macro & ODENet & \cite{hu2022revealing}\\ \cline{1-4}
  \multirow{4}{*}{Multi-scale simulation}
    & \multirow{2}{*}{Micro2macro}
    & RNN & \cite{zhu2021mixseq}\\ \cline{3-4}
    & & GP & \cite{rocha2021fly}\\ \cline{2-4}
    & \multirow{2}{*}{Mixed}
    & GNN & \cite{lino2021simulating}\\ \cline{3-4}
    & & ODENet & \cite{yang2020machine}\\ \cline{1-4}
  \end{tabular}
  \end{center}
  \caption{Data-driven modeling and simulation methods}\label{Tab:data-driven}
\end{table*}

\begin{packed_itemize}
\item \textbf{RNN}: RNN is dedicated to capturing the temporal dependencies, by learning the conditional distributions for current and future temporal status based on the past time series. The MixSeq~\cite{zhu2021mixseq} forecasts the macroscopic time series by aggregating a mixture of microscopic temporal sequences predicted by the seq2seq model. The authors assume that several successive microscopic time series follow a mixture of distributions. Hence cluster them into different components before modeling each component with a seq2seq model configured by specific parameters. It is theoretically shown that with the appropriate estimation for each component, the estimation for the macroscopic time series is enhanced. The experimental results show that MixSeq performs well on both synthetic and real data. The AI-ATS algorithm proposed in~\cite{han2021artificial} utilizes LSTM and GRU to learn the optimal time step lengths and step jumps for the discrete simulation of platelet dynamics in shear blood flow. By adaptively choosing the simulating time step size and jump, AI-ATS reduces the unnecessary redundant calculations for efficient computation and attains comparable accuracy performance to the standard algorithm. In~\cite{ghavamian2019accelerating}, the authors replace the micro model in a finite element simulation with an RNN framework for acceleration in materials. By sampling the results generated by the finite element method, they train the RNN model to understand the historic status, which resembles a nonlinear finite element analysis. Experiments reveal the qualification of RNN as a surrogate.

\item \textbf{CNN}: For the simulations considering spatial characteristics and correlations, the CNN is regarded as a candidate for its capability of describing spatial dependencies. RCNN is proposed in~\cite{guo2021electromagnetic} for Computational EM (CEM) to play as a replacement for the Finite Difference Time Domain (FDTD) method. It takes both RNN and CNN to imitate the complex computations along time frames and space grids for electromagnetic field simulation in a microscopic view.

\item \textbf{GNN}: The GNN is suitable for representing a set of entities, their attributes, and their relationships with each other. The MultiScaleGNN~\cite{lino2021simulating} treats the concerned physical domain as an unstructured set of nodes, and edges are the distance between each pair of nodes. By establishing several graphs corresponding to different scales of spatial resolution, and realizing the message passing forwardly and inversely between these graphs, MultiScaleGNN is able to perform as a multi-scale simulator for continuum mechanics, which is tested on the advection problems and the incompressible fluid dynamics.

\item \textbf{ODENet}: The Ordinary Differential Equations Network (ODENet)~\cite{chen2018neural} is regarded as a replacement of ResNet~\cite{he2016deep}, which depends on the neural network to represent the derivative of variables. Furthermore, it is capable of modeling the time series with irregular time intervals, showing its superiority over RNN. The authors in~\cite{yang2020machine} utilize the ODENet to take place of the chemical master equations for releasing the burden on both memory and computation costs in the single proliferative compartment model. Moreover, the ODENet executes the model reduction in a gene network with autoregulatory negative feedback through multi-time scales. In~\cite{hu2022revealing}, the ODENet is considered for modeling both the classical Lokta-Volterra equations in ecology and the chaotic Lorenz equations in atmospheric turbulence. It embeds the hidden dynamics buried within the massive time sequential data by integrating both machine learning algorithms and ODE modeling.

\item \textbf{GP}: As a typical probabilistic machine learning technique, GP attempts to formulate the observed data with a stochastic process based on Gaussian distribution. When analyzing the stress distribution of a plastic strain band loaded with a transverse tension, the GP method replaces the nested micro-models realized by the finite element approach~\cite{rocha2021fly}. It builds a reduction framework for accelerating the computation on a microscale before integrating for macroscopic analysis.
\end{packed_itemize}

In order to better understand the relationships of the mentioned methods with the simulation scales, we further provide a categorization as listed in Table~\ref{Tab:data-driven}. Though only several representative works are introduced, it indicates the feasibility and appropriateness of utilizing the data-driven methods in the multi-scale simulation. To sum up, both RNN and ODENet are suitable for temporal simulation, while ODENet is able to handle irregular time interval cases. CNN can deal with the simulation task from the spatial perspective. GNN is a wise choice when the concerned simulation problem can be treated as a graph. GP represents the condition when the simulated solution can be fitted by the stochastic distribution. The data-driven methods accelerate the computational process. However, it depends on obtaining the training data as a prerequisite.

\subsubsection{Data-driven multi-scale interaction mechanisms}
Besides the utilization of data-driven methods in the simulation of single-scale and multi-scale physical problems, the ways in which the data-driven methods bridge and integrate different scales are discussed here. We list some representative works in Table~\ref{Tab:data-driven interaction} for a clear view.

\begin{table*}[ht]
  \begin{center}
  \begin{tabular}{c|l|l|l}
  \hline
  \textbf{Category} & \textbf{Scale} & \textbf{Method} & \textbf{Papers}\\
  \hline\hline
  \multirow{6}{*}{Cross-scale interaction mechanisms}
   & \multirow{2}{*}{Micro2macro}
   & GP & \cite{peirlinck2019using}\\ \cline{3-4}
   & & DNN & \cite{wang2020deep}\\ \cline{2-4}
   & Macro2micro & VAE & \cite{bhatia2021machine, ingolfsson2022machine}\\ \cline{2-4}
   & \multirow{3}{*}{Mixed}
   & BN & \cite{hall2021ginns}\\ \cline{3-4}
   & & PCA & \cite{louison2021glimps}\\ \cline{3-4}
   & & RL & \cite{alshehri2022machine}\\ \cline{1-4}
  \end{tabular}
  \end{center}
  \caption{Data-driven multi-scale interaction mechanisms}\label{Tab:data-driven interaction}
\end{table*}

Peirlinck et al.~\cite{peirlinck2019using} model the heart failure process of pigs from multi-scale perspectives. It propagates the measured uncertainties from the cellular level to the organ level by training a Gaussian process regression. To be specific, it decomposes the heart volume analysis into longitudinal growth monitoring, and correlates such longitudinal growth with the measured cellular level uncertainties via the Gaussian process regression.
Wang et al.~\cite{wang2020deep} first design a reduced-order model for generating coarse-grained simulation data in flow dynamics. Then, they take advantage of the Deep Neural Network (DNN) to learn the correlation between the coarse data and the observed fine-scale data.
GINN~\cite{hall2021ginns} is a hybrid approach that combines deep learning with Probabilistic Graphical Models (PGM). It leverages the Bayesian Network (BN), one of the typical PGMs to describe the conditional relationships for key variables across multi-scale in the supercapacitor dynamics. With the data generated by BN, a DNN is developed to establish the correlation between inputs and outputs.
~\cite{alshehri2022machine} introduces the machine learning applications in Computational Molecular Design (CMD) from four aspects: property estimation, catalysis, synthesis planning, and design methods. It creatively proposes the usage of Reinforcement Learning (RL) to link the microscopic molecular design with the macroscopic product characteristics. Concretely, the approach selects the candidate design based on a generative model. On passing the decision to a physical simulation system, the current state and reward are produced, which further guides the evolution direction for the RL algorithm.
GLIMPS~\cite{louison2021glimps} implements the resolution transformation in molecular models based on the Principle Component Analysis (PCA). It captures the structural characteristics of both low-resolution and high-resolution data, and employs the General Linear Model (GLM) to link them.
In~\cite{bhatia2021machine}, a dynamic-importance sampling framework is proposed for bridging the macro and micro scales. It explores the phase space of a macro model, then leverages the Variational Autoencoder (VAE) to project the data onto a latent space, and finally matches the simulations on the microscale. Furthermore, it creates a self-healing mechanism for adjusting the parameters of the macro model inspired by the micro-scale simulation.
~\cite{ingolfsson2022machine} follows a similar technical route as that mentioned in~\cite{bhatia2021machine} to reveal the dynamics of RAS signaling proteins. The scale bridging is realized by mapping the lipid composition into a latent space, to instantiate the molecular simulations on microscales. There is also a step for refining the macro model parameters based on the feedback.

In summary, the data-driven methods are capable of linking the micro and macro scales. Approaches like GP and BN establish the bridge crossing the scales based on the stochastic distribution or probability theory. PCA and VAE realize the dimensional reduction for matching the feature space from high dimension to low dimension. DNN links the multi-scale simulation via a neural network, while RL observes the current system state and reward on the level of one scale, and provides the action based on the design from another scale. The data-driven methods, mainly supported by machine learning techniques, manage to correlate features across the scales as a result of the ability to construct nonlinear relationships.

\subsection{Data and knowledge jointly driven method}
In the domain of multi-scale simulation, for the purposes of high efficiency and fidelity, the data-driven and knowledge-driven methods are considered together, taking advantage of both sides.

\begin{table*}[ht]
  \begin{center}
  \begin{tabular}{l|l|l}
  \hline
  \textbf{Micro-scale simulation} & \textbf{Macro-scale simulation} & \textbf{Papers}\\
  \hline\hline
   Data-driven & Knowledge-driven & \cite{peirlinck2019using, wu2021machine}\\ \hline
   Knowledge-driven & Data-driven & \cite{alshehri2022machine, liu2019exploring}\\ \hline
   Mixed & Mixed & \cite{leung2022nh}\\ \hline
  \end{tabular}
  \end{center}
  \caption{Data and knowledge jointly driven methods}\label{Tab:joint-driven}
\end{table*}

The characterization and analysis of the heart failure procedure are conducted in~\cite{peirlinck2019using}. It models the growth of heart by a stretch-driven growth method, and quantifies the uncertainties via machine learning approaches. The measurements are carried out on the hearts of six pigs. With the help of Bayesian inference, the authors predict the uncertainty probability during the cardiac growth process.
NH-PINN, which is proposed in~\cite{leung2022nh}, converts the multi-scale equation into a homogenized one merging the multi-scale properties. By utilizing the Physics-Informed Neural Network (PINN) for sequentially solving the cell problems and the homogenized equation, NH-PINN manages to improve the accuracy for solving multi-scale problems.
In~\cite{alshehri2022machine}, the authors propose a framework based on RL for molecular design. The current decision is determined by the RL algorithm, while the state and reward are generated through a knowledge-based physical simulation system.
To tackle the general 3D problems with arbitrary material and geometric non-linearity, ~\cite{liu2019exploring} proposes a two-phase framework based on the Deep Material Network (DMN). It first derives the experimental data for training and testing by Direct Numerical Simulation (DNS) tools in the offline stage. Later, it trains the DMN model to learn the relations between the microscale nonlinear properties and the overall macroscale variables, for the purposes of material design and concurrent simulations.
In the field of biology, the authors in~\cite{wu2021machine} develop a framework that combines knowledge and data-driven modeling in the multi-scale analysis of the reconstruction of the sheep tibia over a period of 12 months. Specifically, the overall model consists of the Finite Element (FE) analyzer at the macroscopic level, along with one neural network and another series of neural networks performing as the replacements at the microscopic level. The results reveal a comparative performance of the proposed model in accuracy, while a superb performance in efficiency.

The import of the data-driven methods into the physical knowledge dominant simulation is a seamless combination. It not only accelerates the simulation efficiency by learning the relations hidden in the data, but also guarantees the simulation fidelity by obeying the knowledge rules. We believe that the data and knowledge jointly driven simulation method will attract more and more attention, with applications in multiple research areas.

\subsection{Method Comparison}

In this section, we further discuss the limitations of methods of multi-scale modeling and simulation. To better compare these methods, we further divide the knowledge-driven methods into two categories, i.e.,  mechanistic models and phenomenological models~\cite{abatzoglou1992phenomenological,daniels2015automated}. The mechanistic model, usually established at micro scales, aims to describe the fundamental principles of the system to facilitate first-principle simulation. Differently, phenomenological models are usually established at macro scales and focus on capturing observed trends, patterns, or behaviors of the system~\cite{daniels2015automated}. As the difference compared in Table~\ref{Tab:methodcompar}. We can observe that data-driven methods have relatively smaller computational complexity compared with most knowledge-driven methods, but they have higher overfitting risks and lower explainability. On the other hand, phenomenological knowledge-drive models have smaller computational complexity and smaller overfitting risk compared with mechanistic knowledge-drive models, but they have the lowest modeling ability, resulting in an incapacity to capture numerous details of the system. Altogether, by incorporating both knowledge and data, the data and knowledge jointly driven methods aim to simulate the complex system with low computational complexity, low overfitting risk, high explainability, and high modeling ability.

\begin{table*}[ht]
  \begin{center}
  \resizebox{1\textwidth}{!}{
  \begin{tabular}{l|l|l|l|l}
  \hline
  \makecell[c]{\textbf{Categories of} \\ \textbf{Methods}} & \makecell[c]{\textbf{Computational} \\ \textbf{Complexity}}& \makecell[c]{\textbf{Overfitting} \\ \textbf{Risk}} & \textbf{Explainability}&\makecell[c]{\textbf{Modeling} \\ \textbf{Ability}} \\
  \hline\hline
   Knowledge-driven (mechanistic)& High & Medium & High & High \\ \hline
   Knowledge-driven (phenomenological) & Low & Low & High& Low \\ \hline
   Data-driven & Low & High & Low & High \\ \hline
   Data and knowledge jointly driven & Low & Low & High & High \\ \hline
  \end{tabular}
  }
  \end{center}
  \caption{Comparison of different methods of multi-scale modeling and simulation}\label{Tab:methodcompar}
\end{table*} 

\section{Objective-oriented multi-scale simulation method}\label{sec:objmethod}

\subsection{Methods to achieve Objective A  --- Resolving local fine-grained dynamics}

\begin{table*}[ht]
\begin{center}
\scalebox{0.93}{
\begin{tabular}{c|l|l|l}
\hline
\textbf{Categories}                   & \textbf{Method}           & \textbf{Papers}                   & \textbf{Particle/field based}  \\ \hline
\hline
\multirow{2}{*}{Domain decomposition} & knowledge-driven          & \cite{brooks2009charmm}           & Particle \& Field  \\ \cline{2-4} 
                                      & data-driven               & \cite{mosser2017reconstruction,yang2018microstructural,wang2019meta,liu2016self,wang2019coarse}           & Particle \& field  \\ \hline
\multirow{2}{*}{Multigrid method}     & theory-guided             & \cite{gravemeier2010algebraic,dai2019excmg}           & Field  \\ \cline{2-4}
                                      & technique-introduced      & \cite{yushu2020image}             &  Field   \\ \hline
\end{tabular}
}
\end{center}
\caption{Multi-scale modeling and simulation methods to achieve {\em Objective A}}\label{Tab:type-a}
\end{table*}

Simulating microscale dynamics with fine-grained information requires huge computational costs leading to obstacles before utilization. One promising solution is that the fine-grained modeling is only applied to key parts of the system while the coarse-grained modeling is used in the remaining part.
The methods to achieve \textit{Objective A}, which resolves the localized region of interests into microscale for the simulation of fine-grained behaviors, are mainly categorized into two subcategories: \textit{domain decomposition-based methods} and \textit{multigrid-based methods}. The domain decomposition based methods segment the regions of numerical calculations, calculate them separately, and then integrate them~\cite{weinan2011principles}. The multigrid-based methods, which divide the problem to be solved with different scales of grid splitting, focus on eliminating errors and unnecessary computations by interactions between scales~\cite{weinan2011principles}. Recently, data-driven modeling methods, such as deep learning, have been widely adopted to integrate with domain decomposition-based and multigrid-based methods to enhance the modeling capabilities and computational efficiency, which demonstrates the advantages of synergistic modeling of knowledge and data. These works will be highlighted in the following paragraphs.

\textbf{Domain decomposition based methods.} As a classical multi-scale modeling schema, domain decomposition has played a great role in chemistry~\cite{brooks2009charmm}, biology~\cite{yang2018microstructural}, and physics~\cite{liu2016self} fields. A widely accepted strategy is to use fine-grained accurate models in the local regions of interest and the coarse-grained representation in other regions~\cite{fish2021mesoscopic}. CHARMM~\cite{brooks2009charmm} is a multi-scale molecular simulation platform on which multi-field computational simulation experiments can be performed. The quantum mechanics/molecular mechanics (QM/MM)  model, which applies the QM model to the region of interest and the MM model to the remaining region, can achieve precision and efficiency at the same time. 
Besides, there are many data-driven methods, including variational autoencoder (VAE)~\cite{kingma2013auto}, generative adversarial networks (GAN)~\cite{goodfellow2020generative}, clustering~\cite{xu2005survey} and deep reinforcement learning (DRL)~\cite{mnih2013playing}, utilized to fit the distribution of microscale configuration based on macroscale quantities~\cite{mosser2017reconstruction,yang2018microstructural,wang2019meta,liu2016self,wang2019coarse} in order to handle the heterogeneity of the complex systems. Since recurrent neural networks (RNNs)~\cite{mikolov2010recurrent} have the capability of modeling temporal evolution, Wu et al.~\cite{wu2020recurrent} also model the state changes of microscopic materials affected by history and the macroscale background field using RNNs.

\textbf{Multigrid-based methods.} The multigrid method is an effective computational solution for PDEs (partial differential equations) thanks to its iterative computation and the multi-scale property, which leads to the popular utilization in complex system simulation~\cite{miehe2007multiscale}. 
In the study of turbulent flow simulation, scale transfer operators implemented through plain aggregation algebraic multigrid method could improve the computational efficiency and accuracy~\cite{gravemeier2010algebraic}.
Yushu et at.~\cite{yushu2020image} use image processing techniques in the multigrid method to improve the computational speed of its modeling multi-scale complex systems. 
Dai et al.~\cite{dai2019excmg} propose that the multi-scale multigrid method is a high-accuracy solution for PDEs and design a cascadic multigrid method to further enhance the efficiency.

\subsection{Methods to achieve Objective B --- Completing macroscale mechanisms}

\begin{table*}[ht]
\begin{center}
\scalebox{0.93}{
\begin{tabular}{c|l|l|l}
\hline
\textbf{Categories}                                    & \textbf{Method}           & \textbf{Papers}   & \textbf{Particle/field based}  \\ \hline
\hline
\multirow{2}{*}{Knowledge-driven}                      & FEA                       & \cite{sridhar2016homogenization,ozdemir2008computational}           & Fields \\ \cline{2-4} 
                                                       & REV                       & \cite{yuan2009multiple}                                             & Field \\ \hline
\multirow{2}{*}{Data-driven}                           & NN surrogated DFT         & \cite{allam2018application}                                         & Field \\ \cline{2-4}
                                                       & NN surrogated FEA         & \cite{le2015computational,xie2018crystal}                           & Field \\ \hline
\multirow{2}{*}{Knowledge and data jointly driven}     & NN enhanced FEA           & \cite{hambli2011multiscale}                                         & Field \\ \cline{2-4}
                                                       & FEA guided NN             & \cite{oishi2017computational}                                       & Field \\ \hline
\end{tabular}
}
\end{center}
\caption{Multi-scale modeling and simulation methods to achieve {\em Objective B}}\label{Tab:typeb}
\end{table*}

As mentioned before, the macroscopic simulation on coarse spatio-temporal granularity usually has the drawback of low accuracy due to the lack of microscopic fine information. In particular, it is often difficult to obtain high-precision solutions when encountering systems with strong heterogeneity. The \textit{Objective B} proposed in this literature brings clean fine-grained information to macroscale modeling to promote the simulation, which can be classified into three subcategories: knowledge-driven, data-driven as well as knowledge and data jointly driven methods.

\textbf{Knowledge-driven methods.} The most commonly used knowledge-driven methods include finite element analysis (FEA)~\cite{hughes2012finite} and representative elementary volume (REV)~\cite{hill1963elastic}.
Sridhar et al. \cite{sridhar2016homogenization} use microdynamics differential equations to model the microscopic local resonance phenomena, which is integrated with the macroscale model for continuum materials.
Özdemir et al. \cite{ozdemir2008computational} utilize FEA to model local microstructural heterogeneities and thermal anisotropy with the continuum macroscale.
Yuan et al. \cite{yuan2009multiple} introduce discretized eigendeformation fields at different scales to improve the homogenization method, in which the elasticity solutions of unit cells are precomputed and aggregated to higher scales.

\textbf{Data-driven methods.} In data-driven methods, machine learning models have replaced traditional methods with high computational costs. Density functional theory (DFT), which models the quantum mechanism numerically with heavy computation, is replaced by neural networks in \cite{allam2018application}.
In \cite{le2015computational}, the neural networks take place the FEA to approximate the surface response and to compute the macroscopic stress and tangent tensor components.
Xie et al. \cite{xie2018crystal} utilize graph neural networks, which take the microscale connection of atoms in the crystal and output the material properties, to accelerate the computation.

\textbf{Knowledge and data jointly driven methods.} Recently, physical knowledge and machine learning models have been combined to model complex systems~\cite{karniadakis2021physics}. The deep material network~\cite{hambli2011multiscale} is typical of combining neural networks and REV to model material heterogeneity. The FEA and neural networks are integrated together to model the multi-scale process of bone remodeling. Oishi et al. \cite{oishi2017computational} extract the inherent regularity in a computational mechanics application through machine learning models, and the regularity is then leveraged to design a novel FEA method.

\subsection{Methods to achieve Objective C --- Co-simulating microscale and macroscale dynamics}

\begin{table*}[ht]
\begin{center}
\scalebox{1}{
\begin{tabular}{c|l|l|l}
\hline
\textbf{Categories}                                    & \textbf{Method}           & \textbf{Papers}                                    & \textbf{Particle/field Based}       \\ \hline
\hline
\multirow{2}{*}{Knowledge-driven}                      & Boundary-constraint       & \cite{chakraborty2018hyperdynamics}                & Particle \& field \\ \cline{2-4} 
                                                       & Overlap interaction       & \cite{tao2010nonintrusive,guo2016multi}            & Field \\ \hline
\multirow{2}{*}{Data-driven}                           & Sampling-based            & \cite{ingolfsson2022machine,bhatia2021machine}     & Particle \& field   \\ \cline{2-4}
                                                       & Fitting-based             & \cite{han2017deep,fortunato2022multiscale,lino2022towards,lino2022remus,lino2021simulating}           &   Particle \& field                   \\ \hline
\end{tabular}
}
\end{center}
\caption{Multi-scale modeling and simulation methods to achieve {\em Objective C}}\label{Tab:typec}
\end{table*}

In some scenarios, both the computational efficiency and fine granularity of simulation are important in studying the whole system. However, fine-grained simulation will inevitably have a computationally intensive limitation due to the high level of information granularity~\cite{di2019massively}. Therefore, works~\cite{ingolfsson2022machine,bhatia2021machine,guo2016multi,chakraborty2018hyperdynamics,han2017deep,tao2010nonintrusive,di2019massively} develop techniques to exploit multi-scale modeling and specially designed interactions between scales to achieve both fine-grained and efficient simulation. Consistent with the overall review, we grouped this literature into two categories, i.e., knowledge-driven and data-driven.

\textbf{Knowledge-driven methods.} Traditionally, molecular dynamics (MD) is applied as a high-precision simulation method at the atomic level and is used as the basis for fine-grained simulation in many applications. Coarse-grained simulations, on the other hand, usually focus on different spatio-temporal scales in different tasks. Examples include the lipid scale for RAS protein simulations and the continuum scale for heterogeneous material simulations. In order to achieve the numerical integration effect with high accuracy, Tao et al.~\cite{tao2010nonintrusive} propose the two scale flow method combining fast and slow variables. For modeling the crack propagation of heterogeneous materials, Chakraborty et al.~\cite{chakraborty2018hyperdynamics} design a physics-based continuum crack evolution model called concurrent atomistic-continuum computational model, which consists of FEM and MD. The MD is improved through a novel strain-boost hyperdynamics accelerated time marching scheme to bridge the time gap between different scales. In the electron flow simulation, the Navier-Stokes (NS) equations have been widely used in classic continuum scales while the MD is applied in the microscale for more accurate modeling~\cite{guo2016multi}. In \cite{guo2016multi}, Guo et al. divide the simulation into three parts: the near-wall part, the bulk part, and the overlap part. The near-wall part and the bulk part use the NS equations and MD separately. The overlap part models the coupling between the remaining two parts with a stochastic Eulerian-Lagrangian method.

\textbf{Data-driven methods.}
There are several works~\cite{ingolfsson2022machine,bhatia2021machine,han2017deep} introducing data-driven methods to enhance the coupling of scales in multi-scale modeling, where the main purpose is to more efficiently interact between scales. Bhatia et al.~\cite{ingolfsson2022machine} and Helgi et al.~\cite{bhatia2021machine} adopt variational auto-encoder to embed a latent space for different regions for phase scale in bioprotein simulation and utilize a novel sampling method to select important regions for fine-grained simulation via MD. Recently, Deep Potential~\cite{han2017deep} leverages a neural network to model the potential surface of atoms in a molecular dynamics system, which accelerates the acquisition of atomic potential energy and achieves the quantum mechanical level of accuracy. With the development of the utilization of GNN in atomic simulation, Lino et al. integrate the idea of multi-scale in simulating the continuum mechanics through specially designed GNN~\cite{lino2021simulating}, and there are works follow this novelty~\cite{fortunato2022multiscale,lino2022towards,lino2022remus}.

\subsection{Methods to achieve Objective D --- Discover unknown scales}

Methods to achieve {\em Objective D} reduce high-dimensional complex systems to scales that are easier to understand and investigate. We introduce existing methods to solve the two subproblems of {\em Objective D}, respectively.

\textbf{Scales with linear properties.} High-dimensional complex systems are highly intractable. To address the problem, a series of methods are proposed to obtain simple scales of complex systems. Among them, a typical method is dynamic mode decomposition (DMD)~\cite{schmid2010dynamic,kutz2016dynamic,proctor2016dynamic,tu2013dynamic}, which is a data-driven dimensionality reduction algorithm proposed by Peter Schmid~\cite{schmid2010dynamic}. DMD processes time series data by computing a set of dynamic models that are separately associated with a fixed oscillation frequency and decay/growth rate. The linear treatment largely reduces the dimension of complex systems without the loss of accuracy, and DMD models are widely applied in spatiotemporal fluid flow analysis~\cite{schmid2011applications,hemati2014dynamic,habibi2020data}. 

\textbf{Scales with preserved relevant information.}  Researchers seek to identify causal emergence for multi-scale complex systems. For example, Griebenow et al.~\cite{griebenow2019finding} utilize spectral clustering to find the right scale of a complex network with causal emergence. Researchers also discover that macro scales defined by coarse graining the microscale elements show surprising causal emergence~\cite{hoel2013quantifying}. By defining and evaluating the metric of effective information (EI) both at the micro and macro scales, causal interactions are observed to peak at a macroscale. The coarse scales with higher causalities facilitate investigations into complex systems more effectively. Data-driven models~\cite{bar2019learning,lee2020coarse} are proposed to identify coarse-scale partial differential equations (PDEs) that describe the system evolution at the macroscopic level. A series methods to identify slow variables in dynamical complex systems~\cite{perez2013identification,singer2009detecting,noe2013variational,wehmeyer2018time,bramburger2020sparse} are proposed, such as variational approach~\cite{noe2013variational}, autoencoders~\cite{wehmeyer2018time,li2023learning}, diffusion maps~\cite{singer2009detecting}. Meanwhile, governing equations~\cite{champion2019data,brunton2016discovering} or common principles~\cite{huang2018mesoscience,chen2022automated} of the system are extracted. The discovery of governing equations and common principles is essential in data-rich fields that lack well-established theories. Recently, the renormalization group (RG) method has shown great potential to deal with high-dimensional systems~\cite{koch2018mutual,li2018neural,di2022deep}. The RG method is intimately related to scale invariance and conformal invariance, symmetries in which a system appears the same at all scales, i.e., so-called self-similarity.

\begin{table}[]
\centering
\setlength{\leftmargini}{0.4cm}
\begin{tabular}{m{4cm} | m{5cm} | m{4cm} }
\hline
\textbf{Category} & \textbf{Method} & \textbf{Examples} \\ \hline\hline
Scale with linear properties & Dynamic mode decomposition~\cite{schmid2010dynamic,kutz2016dynamic,proctor2016dynamic,tu2013dynamic} & ~\cite{schmid2011applications,hemati2014dynamic,habibi2020data} \\ \hline
\multirow{4}{*}{\makecell[c]{Scale with preserved \\ relevant information}}
& Causal emergence & \cite{griebenow2019finding,hoel2013quantifying} \\ \cline{2-3} 
 & Coarse-scale partial differential equations & \cite{bar2019learning,lee2020coarse} \\ \cline{2-3} 
 & Slow variables & \cite{perez2013identification,singer2009detecting,noe2013variational,wehmeyer2018time,bramburger2020sparse,li2023learning}\\ \cline{2-3} 
 & Governing equations  & \cite{champion2019data, brunton2016discovering, huang2018mesoscience, chen2022automated} \\ \cline{2-3}
 & Renormalization group &  \cite{koch2018mutual,li2018neural,di2022deep} \\ \hline
\end{tabular}%
\caption{Multi-scale modeling and simulation methods to achieve {\em Objective D}}
\end{table}

\subsection{Methods to achieve Objective E  --- Learning dynamics at the discovered scale}

Methods to achieve {\em Objective E} solve dynamical complex systems without clear scales by learning effective dynamics. To deal with the challenge of high dimensions, there are two types of methods to learn effective dynamics at a lower-dimensional scale, which can be further utilized to model dynamical systems.

\textbf{Knowledge-driven methods.}
Knowledge-driven methods leverage mathematical models to learn dynamical complex systems. The widely used approaches include wavelet-based methods~\cite{mallat1989theory,beylkin1998multiresolution,gilbert1998comparison,mehraeen2006wavelet}, multi-scale homogenization~\cite{chen2006generalized,yu2002multiscale,li2008generalized}, and stochastic methods~\cite{fish2011nonintrusive,hu2017adaptive}. For example, the wavelet representation~\cite{mallat1989theory},  which is an orthogonal multiresolution representation computed by a pyramidal algorithm, can be applied to various scenarios of scale compression.   To deal with large spaces, Jacob et al.~\cite{fish2011nonintrusive} propose a reduced-order homogenization method combined with the Karhunen–Loeve expansion and stochastic collocation method. Molei et al.~\cite{tao2010nonintrusive} also develop a new class of multi-scale integrators for stiff ODEs and SDEs for Hamiltonian systems with hidden dynamics.

\textbf{Data-driven methods.} Data-driven methods leverage machine learning or deep learning models to learn the effective dynamics of complex systems. Vlachas et al.~\cite{vlachas2021accelerated} accelerate simulations of molecular systems by learning dynamics at the coarse scale modeled by autoencoders and applying a probabilistic mapping between coarse and fine scales via a mixture density network. Some other studies also use autoencoders to model dynamical systems~\cite{vlachas2022multiscale,lee2020model,fukami2020convolutional}.  Besides, variational autoencoder (VAE) can be utilized to learn latent-space dynamics of high-dimensional systems~\cite{hernandez2018variational}. This is achieved by reducing complex, nonlinear processes to a single embedding with high fidelity. The performance of long short-term memory networks (LSTMs)~\cite{hochreiter1997long} and neural ordinary differential equations (NODEs)~\cite{chen2018neural} are explored in learning latent-space representations of dynamical equations~\cite{maulik2020time,maulik2021reduced}. 
Deep Koopman neural networks~\cite{brunton2021modern} and LSTM are utilized to model the dynamics of slow variables and fast variables, respectively~\cite{li2023learning}.
In addition, sampling strategies~\cite{champion2019discovery,bhatia2021machine,gong2021advanced,dupuis2012importance,manohar2019optimized} are proposed for modeling multi-scale dynamics. For example, Harsh et al.~\cite{bhatia2021machine} present a dynamic-importance sampling approach for adaptive multi-scale simulations. It uses machine learning to dynamically sample the phase space by a macro model using microscale simulations, which enables automatic feedback from the micro to the macro scale.

\begin{table}[]
\centering
\setlength{\leftmargini}{0.4cm}
\begin{tabular}{m{4cm} | m{5cm} | m{4cm} }
\hline
\textbf{Category} & \textbf{Method} & \textbf{Examples} \\ \hline\hline
\multirow{3}{*}{Knowledge-driven} & Wavelet-based methods &  \cite{mallat1989theory,beylkin1998multiresolution,gilbert1998comparison,mehraeen2006wavelet}\\ \cline{2-3}
& Multiscale homogenization & \cite{chen2006generalized,yu2002multiscale,li2008generalized}\\ \cline{2-3}
& Stochastic methods & \cite{chen2006generalized,yu2002multiscale,li2008generalized}\\ \hline

\multirow{3}{*}{\makecell[c]{Data-driven}}
& Autoencoders & \cite{vlachas2021accelerated,vlachas2022multiscale,lee2020model,fukami2020convolutional,hernandez2018variational} \\ \cline{2-3} 
 & Neural ODEs & \cite{chen2018neural,maulik2020time,maulik2021reduced}\\ \cline{2-3}
 & Deep Koopman neural networks & \cite{li2023learning}\\ \cline{2-3}
 & Sampling strategies & \cite{champion2019discovery,bhatia2021machine,gong2021advanced,dupuis2012importance,manohar2019optimized}\\ \hline
\end{tabular}%
\caption{Multi-scale modeling and simulation methods to achieve {\em Objective E}}
\end{table}


\section{Application of multi-scale simulation in typical complex systems}
\subsection{Matter systems}
\subsubsection{Atomic and molecular systems}

Traditionally, Car-Parrinello molecular dynamics (CPMD) is the widely adopted multi-scale simulation method for calculating the configuration of the atomic and molecular system, where the density functional theory (DFT)~\cite{car1985unified} deals with the electronic level quantum mechanics and the classic mechanics are modeling with Newton's laws.
Even though CPMD will bring very high simulation accuracy, the computation will be very large and the utilization is limited. To achieve both effectiveness and efficiency, CHARMM proposed that the local region of interest could be different from the remaining region in simulating methods, which is called quantum mechanics - molecular mechanics (QM-MM)~\cite{brooks2009charmm}. 
Recently, Weinan et al. propose deep potential~\cite{han2017deep}, which introduced deep learning techniques into the complex system simulation, and the effectiveness and efficiency are both obtained at the system level. Following this, GNNs are also introduced to solve the complex problem of turbulence modeling and continuum modeling~\cite{fortunato2022multiscale,lino2022towards,lino2022remus,lino2021simulating}.

\subsubsection{Biomolecular systems}

Multi-scale modeling provides better solutions in many complicated tasks, such as novel protein structure discovery, remodeling of membranes, and biological stress modeling.  
For example, in \cite{bhatia2021machine}, which is aimed at achieving simulation with macroscale space size, macroscale temporal length, and microscale precision, the different scales interact with each other via the consistency at the boundary between different scales. 
Stansfeld et al. design a multi-scale method to simulate the assembly and the interactions of membrane protein/lipid, in which the macroscale is at molecular dynamics while the microscale reaches atomistic resolution~\cite{stansfeld2011coarse}. In the collagen modeling application, Masic et al. reveal how the physics of the microscale at the molecular level contributes to the mechanics at the macro behavior level~\cite{masic2011observations}.
\subsubsection{Chemical systems}

\begin{figure}[t]
	\centering
	\subfigure[]{\label{fig:rules_tran}\includegraphics[width=.18\textwidth]{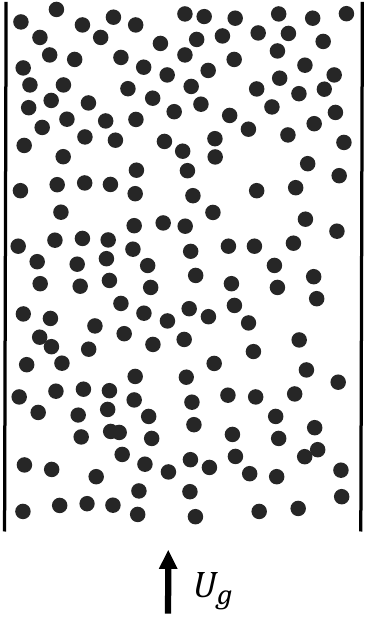}}
	\hspace{-0.1cm}
   \subfigure[]{\label{fig:rules_visit} \includegraphics[width=.18\textwidth]{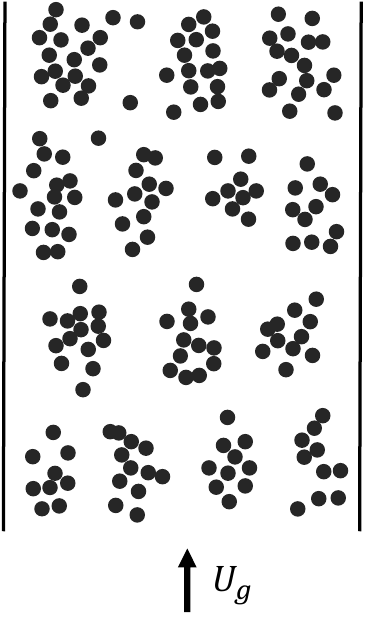}}
	\hspace{-0.1cm}
   \subfigure[]{\label{fig:rules_visit} \includegraphics[width=.18\textwidth]{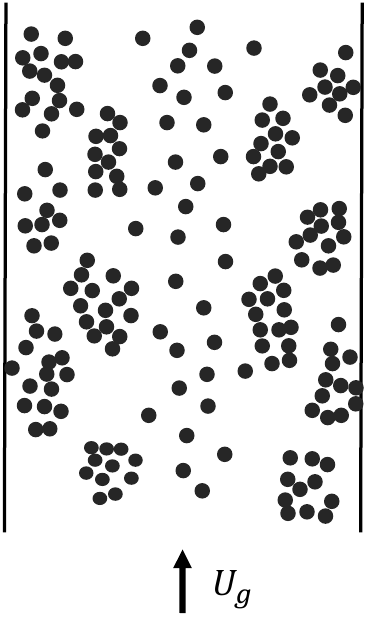}}
   \vspace{-0.2cm}
	\caption{Three different gas–solid two-phase flow with the same amount of solids and the same gas flow rate $U_g$. The different mesoscale structures lead to different properties of the system~\cite{li1994particle}.}\label{fig:chonlyglobal}
\end{figure}

Chemical systems are also a category of systems with multi-scale structures.
For example, a widely existing complex system in nature and engineering with multi-scale structures is the gas–solid two-phase flow, which is a non-equilibrium system with significant heterogeneity.
Specifically, there are two mechanisms in this system, including mechanisms of particles and gas.
Then, in a concurrent-up gas–solid two-phase flow as shown in Figure~\ref{fig:chonlyglobal}, the gas tends to find an upward path with minimum resistance. If the mechanism of gas dominates the systems, which occurs when the flow velocity is particularly large, particles will be separated as much as possible to minimize the resistance to gas, thereby minimizing energy consumption for transporting and suspending particles per unit volume of gas $W_{st}$.
On the other hand, the particles tend to minimize their potential energy.
If the mechanism of particles dominates the systems, which occurs when the flow velocity is particularly small, particles will be as concentrated as possible to minimize their potential energy, thereby minimizing the voidage $\epsilon$ of the gas–solid flow.
However, in most scenarios in nature and engineering, the flow velocity is neither too large nor too small, and neither mechanism is able to dominate the system.
Thus, two mechanisms co-exist and compete with each other in the system, which causes a compromise in reaching the optimal states corresponding to the two mechanisms, i.e., minimal $W_{st}$ and minimal $\epsilon$, respectively.
During the competition of the two principles of minimum potential energy for particles and minimum resistance to gas, the mesoscale structure is formed, i.e., dense particle clusters surrounded by the dilute broth, which are denoted by dense phase and dilute phase, respectively.

From the microscale, by only observing the interaction between each particle and its surrounding circular area, we can only observe two different mechanisms corresponding to the dense phase and the dilute phase.
From the macro scale, only the global state variables cannot fully characterize the system. For example, as we can observe from Figure~\ref{fig:chonlyglobal}, the three systems have the same global state variables including the same gas flow rate $U_g$, and the amount of solids.
The different mesoscale particle cluster sizes lead to the difference between Figure~\ref{fig:chonlyglobal}(a) and (b); The existence of a core-annulus flow further leads to the difference between Figure~\ref{fig:chonlyglobal}(b) and (c). 
Thus, the mesoscale structure must be observed and modeled to fully characterize the complex system.

How to model the mesoscale structure of the gas–solid two-phase flow has been studied in numerous existing studies. 
\citeauthor{li1994particle} et al.~\cite{li1994particle} propose the energy-minimization multi-scale model (EMMS) model.
In this model, the Gas–solid two-phase flow is described by eight variables, of which the most important variables include the fraction of dense phase $f$, the voidage of the dense phase $\epsilon_c$, the voidage of the dilute phase $\epsilon_d$, and the diameter of particle clusters $d_{cl}$, etc. 
They find that instead of minimizing the energy consumption for transporting and suspending particles per unit volume of gas $W_{st}$ or the voidage $\epsilon$ in the system, the energy consumption for the suspension and transportation of particles with respect to unit mass of particles $N_{st}$ is minimized.
Further, \citeauthor{ge2002physical} et al.~\cite{ge2002physical} propose a numerical algorithm to solve this problem.
Specifically, this algorithm defines a two-dimensional solution space of $(\epsilon_c,\epsilon_d)$. For each pair of $(\epsilon_c,\epsilon_d)$, this algorithm searches for the value of $f$ such that $U_{si}$ calculated based on its definition equals $U_{si}$ calculated based the balance conditions including mass balance, momentum balance, and pressure balance, etc.
Specifically, $U_{si}$ represents the slip velocity between two phases.
Then, the problem can be solved by traversing all feasible $(\epsilon_c,\epsilon_d)$ within the solution space and finding the $(\epsilon^*_c,\epsilon^*_d)$ with minimum $W_{st}$.
The EMMS model is further extended to a more general multi-objective variational  methodology~\cite{mo2020analysis, li2018mesoscience}. Based on this methodology, the target system can be mathematically expressed by the multi-object optimization (MOP) problem. 
\citeauthor{mo2020analysis}~\cite{mo2020analysis} validate the consistency of the solutions obtained based on the EMMS model compared with the solutions obtained based on the MOP model through numerical analysis. What's more, the MOP model can further explore those solutions not found by the EMMS model.

\subsection{Social systems}

\subsubsection{Transportation systems}

\begin{table*}
\resizebox{1\textwidth}{!}{
\begin{tabular}{c|p{3cm}|p{2.5cm}|p{2.5cm}|p{2.5cm}|p{2.5cm}}
\hline
\textbf{Categories} & \textbf{Research Problem}       & \textbf{Method}  & \textbf{Macroscopic} & \textbf{Microscopic}        & \textbf{Papers}                                        \\ \hline
\multirow{18}{*}{Spatial}      & How traffic performance can be improved is unclear (Objective C)   & Multi-scale model predictive control approach     & Network level perimeter control    & Local level signal control                                    & \cite{yang2017multi}        \\ \cline{2-6}

        & Unclear the connectivity relationship within a multi-modal urban transportation network (Objective B) & Multi-modal and multi-scale GIS-T data model  & Bus, metro and street transit networks & Bus route segment, metro line, street segments, walking links & \cite{chen2011multi}     \\ \cline{2-6}


         & Unclear changes of future traffic (Objective B)    &  Hierarchical graph convolution network     & Regions     & Road segments    & \cite{guo2021hierarchical} \\ \cline{2-6}

         & Unclear changes of future traffic congestion (Objective C)    &  Cross-scale spatiotemporal GNN     & Regions     & Road segments    & \cite{wang2023contagion} \\ \cline{2-6}

        & Unclear mechanism of the heterogeneous traffic streams (Objective B)   & Mean field games theory     & Network-level traffic streams     & Individual vehicle behavior        & \cite{kachroo2017multiscale}     \\ \hline

\multirow{10}{*}{Temporal}

        & Unclear mechanism of driver behavior  (Objective A)   & A combined CNN-RNN based multi-task learning model  & Long time-scale mental cognitive process      & Short time-scale driver’s physical behaviors     & \cite{xing2021multi}     \\ \cline{2-6}

        & Unclear changes of future traffic (Objective B) & Multi-Step dependency relation model & Short time-scale traffic & Long time-scale traffic & \cite{liu2022msdr} \\ \cline{2-6}

        & Unclear mechanism of urban freight operations  (Objective A)   & Multi-scale agent-based simulation model    & Long-term strategic decision     & Short-term operational decision         & \cite{alho2017multi}     \\ \hline
\end{tabular}
}
\caption{Multi-scale modeling and simulating methods in transportation.}\label{Tab:tranportation_multiscale}
\end{table*}

In the field of transportation, the traffic flow theory is to explore the spatial and temporal patterns of traffic, which has developed several related models. These models can be classified into two groups: microscopic models and macroscopic models. The microscopic models aim to characterize the interactions between the vehicles, while the macroscopic ones focus on the aggregated behaviors of a group of vehicles. Their basis is to investigate the relationship between the density, speed, and flow volume. Specifically, the density $\rho$ is the average number of vehicles passing a unit length of the road segment. The speed  $v$ is the average speed of all the vehicles in a road segment. The flow volume $q$ is the average number of vehicles in a road segment in a time slot. Their relationship can be formulated as follows:
\begin{equation}
q=\rho \cdot v.
\end{equation}

\begin{packed_itemize}
\item \textbf{Macroscopic traffic models}. Such models focus on the macroscopic characteristics of traffic flow, which regards the traffic flow as the continuous fluid composed of a large amount of vehicles. Based on the theory of fluid mechanics, the researchers derive the partial differential equations for the speed and density to characterize the traffic dynamics. Formally, according to the conservation principle of traffic flow, it satisfies the following continuity equation:
\begin{equation}
\frac{\partial (\rho v)}{\partial x} + \frac{\partial v}{\partial t}=0,
\label{equ:lwr}
\end{equation}
where $\rho(x,t)$ and $v(x,t)$ are the flow density and average speed at location $x$ at time $t$, respectively.
This equation is built upon the traffic fluid mechanics model called Lighthill-Whitham-Richards (LWR) model~\cite{lighthill1955kinematic}.
Particularly, for the speed $v(x,t)$, the LWR model assumes that the traffic speed and density satisfy the relationship of $v(x,t)=v_e(\rho(x,t))$. Thus, the Equation~\ref{equ:lwr} can be rewritten as
\begin{equation}
(v_e+\rho \frac{\partial v_e}{\partial \rho})\frac{\partial (\rho v)}{\partial x} + \frac{\partial v}{\partial t}=0.
\label{equ:lwr2}
\end{equation}

The above equation captures the nonlinear traffic features such as the formation and propagation of shock waves in traffic flow, which facilitates to explore various complex traffic phenomena, e.g., the stop-and-go trend in traffic congestion.

\item \textbf{Microscopic traffic models}. These models are to learn the moving states of each vehicle in a certain traffic environment, including car-following models and cellular automaton models.

The car following models study the dynamic process of each vehicle following the leading vehicle based on the assumption that there is no overtaking phenomenon. Traditionally, their general mathematical expression can be shown as follows,
\begin{equation}
\frac{dv_n(t)}{dt}=f(v_n(t),\Delta v_n(t), \Delta x_n(t)),
\end{equation}
where $v(t)$ denotes the speed of vehicle $n$, and $\Delta v_n(t), \Delta x_n(t)$ are the speed difference and displacement difference between vehicle $n$ and its front vehicle. $f(\cdot)$ represents the car following model.
To implement $f(\cdot)$, many researchers propose various methods by modeling drivers' behavior of acceleration or deceleration~\cite{newell1961nonlinear,bando1995dynamical,helbing1998generalized,treiber2000congested,nakayama2001effect,zhao2005new,ge2006effect,xiaomei2007stability,xie2008stabilization,peng2010dynamical}. For example, Reusche and Pipes~\cite{pipes1953operational} proposed the first car following model
\begin{equation}
\frac{dv_n(t+\tau)}{dt}=\lambda \Delta v_n(t),
\end{equation}
which considers the strength of the driver's response to the stimulus he receives from his driving environment. In this equation, $\tau$ is the delayed response time of the vehicle, and $\lambda$ is the strength coefficient of a driver in response to the stimulus.

The cellular automaton models are discrete in both space and time, thus it is suitable for parallel computation to accelerate the simulation of microscopic traffic flow. The simplest cellular automaton model is one-dimensional where a single road segment is divided into discrete regular grids. Each grid represents a cell with the finite states. At any time, the state of a cell is either empty or occupying a vehicle. Further, the state of a cell at the next time is determined by itself and the state of its neighboring cells at the current time. Since a cell cannot have more than one vehicle, we define the following rule: if the cell ahead is empty, the vehicle can move one step forward; otherwise, it has to keep waiting. According to this rule, all vehicles could update their states at the next time. Based on it, several extended models~\cite{helbing1999cellular,chowdhury2000statistical,kerner2002cellular,gao2007cellular,chakroborty2008microscopic,tao2008synchronized} are proposed to capture more realistic phenomena that occur in traffic flows, which helps to better understand the traffic dynamics and implement the effective traffic control.
\end{packed_itemize}

Although we can adopt the microscopic traffic models to infer the dynamics of massive microscopic states in transportation systems, it is not easy to derive the macroscopic structure emerged from these states. One important problem is that we cannot accurately realize the scale at which the macroscopic structure will emerge. This is because the accumulated noise from the microscopic scale will prevent us from identifying the macroscopic structure. To solve it, multi-scale modeling is proposed to fully utilize the advantages of microscopic and macroscopic scales for various traffic applications. Through the reviews of recent works, we classify them into two groups in terms of multi-scale properties, i.e., space and time, as listed in Table~\ref{Tab:tranportation_multiscale}.
\begin{packed_itemize}
\item \textbf{Spatial multi-scale modeling}. The works belonging to this type adopt different scales of space to design the corresponding models. Authors in~\cite{guo2021hierarchical} first used the spectral clustering method to construct the macro graph of regions from the road network, and then proposed the hierarchical graph convolutional network to forecast the traffic by operating both the micro and macro traffic graphs. Yang et al.~\cite{yang2017multi} proposed a model predictive control based approach to model the interaction between the network level control and the local level control to optimize the performance at the local and the network level as a whole. Chen et al.~\cite{chen2011multi} developed a multi-scale and multi-modal Geographical Information Systems for Transportation (GIS-T) data model to implement the integration and multiple representations of the urban transportation networks that include bus route segments, metro lines, street segments, walking links. Authors in~\cite{kachroo2017multiscale} focused on analyzing the heterogeneous traffic streams from connected vehicles and normal vehicles. They proposed a mean field game framework to establish the connection between the individual vehicle behavior and overall macroscopic traffic streams. In this framework, the connected vehicles controlled microscopically can influence and control overall traffic streams.
Recently, a cross-scale spatio-temporal GNN~\cite{wang2023contagion} was proposed to forecast future traffic congestion, as shown in Figure~\ref{fig:transport_example}. This approach highlights the advantages of multi-scale simulation methods in intelligent systems. As illustrated in this study, the complex interplay between microscopic and macroscopic dynamics of traffic congestion significantly impacts traffic congestion. Only modeling one of them cannot capture this complexity, resulting in inaccurate prediction performance. To address this challenge, the authors proposed to model the interaction between macro- and micro-dynamics of congestion, effectively addressing \emph{Objective C}. To be specific, they adopted PINN to extract valuable macroscopic information for guiding the prediction of microscopic traffic congestion. Meanwhile, they designed a micro-macro transformation mechanism to aggregate microscopic states into macroscopic ones in a differentiable manner. This cross-scale modeling significantly improved the prediction performance of traffic congestion.

\begin{figure}[t]
	\centering
    \includegraphics[width=0.6\textwidth]{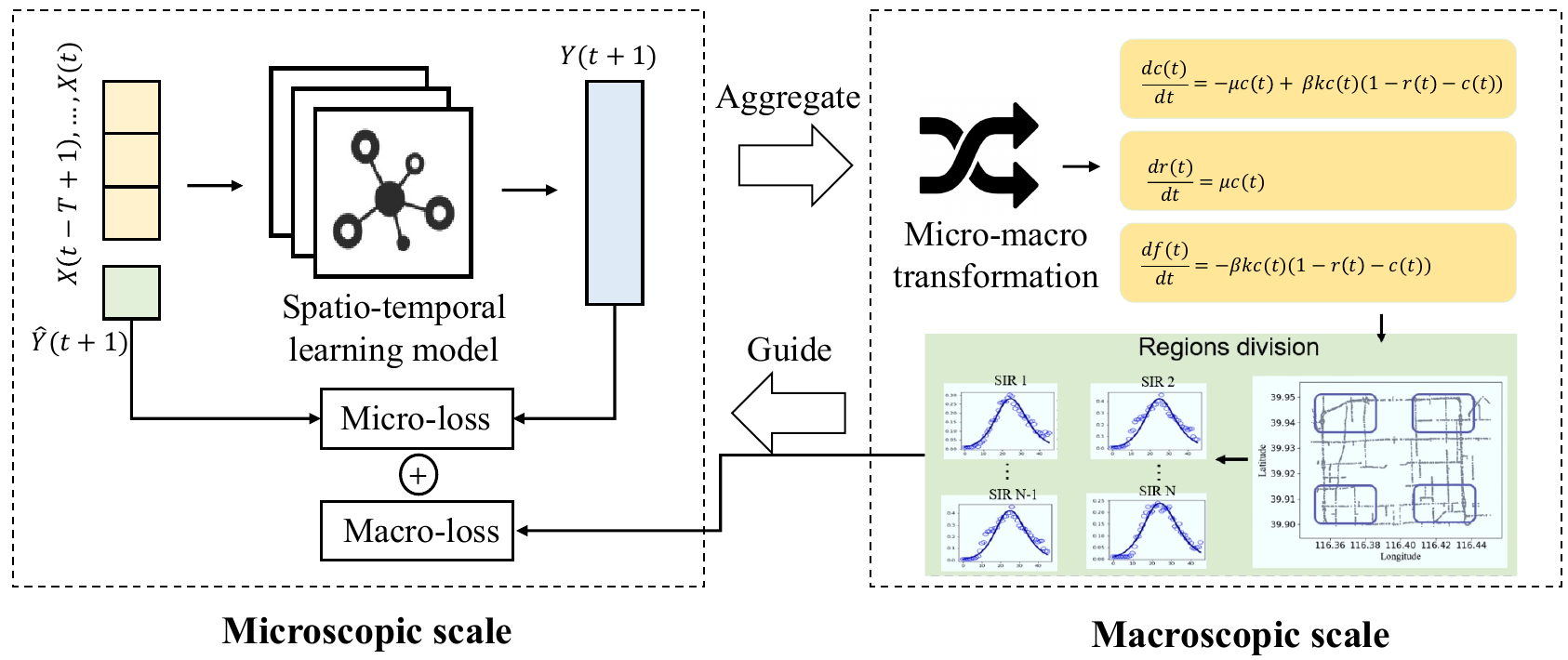}
	\caption{The cross-scale spatio-temporal graph neural network~\cite{wang2023contagion} utilizes multi-scale simulation methods to forecast future traffic congestion (\emph{Objective C}).}
\label{fig:transport_example}
\end{figure}

\item \textbf{Temporal multi-scale modeling}. This type of methods focuses on exploring the advantages of multi-scale modeling in the time dimension. To be specific, Yang et al.~\cite{xing2021multi} proposed to jointly model the driver behaviors in different time scales for behavior reasoning. In this work, the mental cognitive process is recognized to have longer time-scale patterns, while drivers' physical behavior has fast dynamics. A multi-task framework with the CNN-RNN network is developed to process the task of multi-scale driver behavior recognition. Authors in~\cite{liu2022msdr} designed a multi-step dependency relation network to learn the relations between the short-time-scale traffic features and long-time-scale ones. Alho et al.~\cite{alho2017multi} focused on modeling urban freight operations. They proposed a novel agent-based framework integrating the models with different time resolutions, including a long-term strategic decision model and a short-term operational decision model. The former simulates the agent decisions made for one year, and the latter simulates the fast decision-making process. The authors also implemented it on an open-source mobility simulation platform and verified the effectiveness of their proposed approach.
\end{packed_itemize} 

\subsubsection{Epidemic contagion systems}

Accurately modeling the transmission pattern of the epidemic has been a long-standing research topic for decades, which is attracting more and more public attention with the outbreak of COVID-19. The epidemic dynamics is tightly entangled with human mobility \cite{jia2020population,kraemer2020effect,jewell2021s,wu2020nowcasting}, economic development \cite{haw2022optimizing,guan2020global,spelta2020after}, and vaccine distribution \cite{chen2022strategic,rotesi2021national,wagner2021vaccine,yang2022equitable,mcadams2020incentivising}, forming a complex system. In this survey, we focus on the simulation paradigm of epidemic contagion systems of the recent advances, which can be divided into 3 categories as shown in Table~\ref{Tab:epidemic}.

\begin{table*}[ht]
	\begin{center}
	
\resizebox{1\textwidth}{!}{
		\begin{tabular}{c|l|l|l}
			\hline
			\textbf{Categories} & \textbf{Scales} & \textbf{Method} & \textbf{Papers}  \\
			\hline\hline
			\multirow{3}{*}{Single-scale simulation} &
			\multirow{2}{*}{Macroscopic model}
			& Regression model &  \cite{jia2020population,kraemer2020effect,jewell2021s,iacus2020human}  \\ \cline{3-4}
			& & Compartmental model  & \cite{maier2020effective,lai2020effect,ross1916application}    \\ \cline{2-4}
			& Microscopic model & Contact Network  & \cite{murphy2021deep,qian2021scaling,aleta2020modelling,abueg2021modeling,nishi2020network} \\ \cline{1-4}
			\multirow{2}{*}{Multi-scale simulation} &
			\multirow{2}{*}{Multi-scale model}
			& Metapopulation &  \cite{balcan2010modeling,balcan2009multiscale,zhang2017spread,chinazzi2020effect,wu2020nowcasting,zhou2020effects,chang2021mobility,zhang2020changes,brockmann2013hidden}  \\ \cline{3-4}
			& & High-order network  & \cite{iacopini2019simplicial,st2021universal}    \\ \cline{1-4}
		\end{tabular}
		}
	\end{center}
\caption{Simulation methods of epidemic contagion systems}
\label{Tab:epidemic}
\end{table*}

\textbf{Regression model.} For macroscopic simulation, regression models are widely used. During the first strike of the COVID-19 pandemic, researchers use generalized linear model (GLM) \cite{kraemer2020effect}, multiplicative exponential model \cite{jia2020population}, or multilevel Bayesian regression model \cite{jewell2021s} to investigate how human mobility and control measures affect the early transmission of the epidemic. Specifically, the risk source model proposed in \cite{jia2020population} is shown in \eqref{eq:risk_source}:
\begin{equation}
	\label{eq:risk_source}
	y_{i}=c \prod_{j=1}^{m} \mathrm{e}^{\beta_{j} x_{j i}} \mathrm{e}^{\sum_{k=1}^{n} \lambda_{k} I_{i k}},
\end{equation}
where $y_i$ is the number of the cumulative (or daily) confirmed cases in prefecture $i$; $x_{1i}$ is the cumulative population outflow from the source region of the epidemic to prefecture $i$ in different time periods. $x_{2i}$ and $x_{3i}$ are the GDP and the population size of prefecture $i$, respectively. $\lambda_k$ is the fixed effect for province $k$, $n$ is the number of prefectures considered in the analysis, $I_{ik}$ is the dummy for prefecture $i$, where $I_{ik}=1$ if $i\in k$, otherwise $I_{ik}=0$. Finally, $c$ and $\beta_j$ are the free parameters to be calibrated. From this equation, the author accurately models the macroscopic epidemic dynamics. Similarly, the regression model in \cite{jewell2021s} and \cite{kraemer2020effect} are as shown in \eqref{eq:multilevel_bayesian_regression} and \eqref{eq:glm} accordingly. All these methods use a simple regression model to depict the aggregated epidemic outcome ($y_{i, t}$ in \eqref{eq:multilevel_bayesian_regression} and $Y(t)$ in \eqref{eq:glm}) with other variants such as mobility ($M_{i,t}$ in \eqref{eq:multilevel_bayesian_regression} and $M(t-5)$ in \eqref{eq:glm}), population ($X_{i}$ in \eqref{eq:multilevel_bayesian_regression}), the polymerase chain reaction (PCR) test ($IT(t)$ in \eqref{eq:glm}) and temperature ($T_{i,t}$ in \eqref{eq:multilevel_bayesian_regression}).
\begin{equation}
	\label{eq:multilevel_bayesian_regression}
	y_{i, t}=\alpha_{c_{t}}+X_{i} \beta+T_{i, t} \theta+C_{s_{i}, t} \phi+M_{i, t} \gamma_{c_{i}, t}+\epsilon_{i, t},
\end{equation}
\begin{equation}
	\label{eq:glm}
	Y(t) = Y(t-4) + IT(t) + M(t-5) + IM(t).
\end{equation}

{\textbf{Compartmental model.}} Another type of macroscopic model for epidemic modeling is the famous compartmental model, which can be traced to the 20th century \cite{ross1916application}. Classic compartmental models, such as the SEIR model, divide the whole population in a region into different compartments as Susceptible~(S), Exposed~(E), Infected~(I), and Recovered~(R) and use the following ordinary differential equations (ODE) to depict the aggregated transmission process:
\begin{equation}
\begin{aligned}
	{\der S}/{\der t} &= - {\beta SI}/{N}, \\
	{\der E}/{\der t} &= {\beta SI}/{N} - \sigma E, \\
	{\der I}/{\der t} &=  \sigma E - \gamma I,\\
	{\der R}/{\der t} &= \gamma I,
\end{aligned}
\end{equation}
where $S$, $E$, $I$, and $R$ are the susceptible, asymptomatic, infected, and recovered people, respectively.
$N=S+E+I+R$ is the total population of the region. For the parameters, $\beta$ is the infection rate, $\gamma$ is the recovery rate, and $\sigma$ is the incubation period of the disease. Specifically, the basic reproductive number of the disease is $R_0 = \beta / \gamma$, which denotes the expected number of cases directly generated by one case in a population where all individuals are susceptible. When $R_0>1$, the disease can start to transmit in an exponential pattern. When $R_0 < 1$, the disease will not become an epidemic since the infected people will recover before it can transmit to more people. These compartmental models are macroscopic since the variables are the total number of people in different states, where the heterogeneous contact patterns are replaced by a homogeneous mixing assumption. It limits their expressive power since they can only capture the exponential growth pattern, which only exists during the early beginning of the epidemic and quickly changed due to the introduction of non-pharmaceutical interventions \cite{flaxman2020estimating,lai2020effect,snoeijer2021measuring}. 

To extend the expressive power of classic compartmental models, Maier et al. \cite{maier2020effective} introduce a ``containment and quarantine'' mechanism to the classic SIR model as follows:
\begin{equation}
\begin{aligned}
	\partial_{t} S &= -\alpha SI - \kappa_0 S, \\
	\partial_{t} I &= \alpha SI - \beta I - \kappa_0 I - \kappa I, \\
	\partial_{t} R &= \beta I + \kappa_0 S, \\
	\partial_{t} X &= (\kappa + \kappa_0) I,
\end{aligned}
\end{equation}
where $\alpha$ and $\beta$ are the infection rate and the recovery rate accordingly, and the self-containment rate that affects both susceptible and infected people is $\kappa_0$, which assumes the self-containment process will prevent any transmission for these parts of people and directly turn the susceptible and infected people to the new $X$ state (which do not act with any states). Besides, there is also a quarantine rate $\kappa$ that only influences the infected people. By the new mechanism that protects part of the susceptible people, the proposed model successfully extends the expressive power of the SIR model to capture sub-exponential increase patterns.

\textbf{Contact network model.}
For microscopic modeling of the contagion system, most of the research adopts the network approach to represent individual-level contact patterns \cite{murphy2021deep,qian2021scaling,aleta2020modelling,abueg2021modeling,nishi2020network}. Qian et al. \cite{qian2021scaling} use smart card data in urban transit systems to construct individual-level contact networks. Specifically, a classic susceptible-infectious-susceptible (SIS) model is used to model the disease transmission on the network as follows:
\begin{equation}
	p_{i, t}=1+p_{i, t-1}\left(q_{i, t}-r\right)-q_{i, t}, \forall i \in V.
\end{equation}
In the above equation, $p_{i,t}$ represents the probability for node $i$ infected at time $t$, and the recovery rate is noted as $r$. The variable $q_{i,t}$ represents the probability that node $i$ is in the susceptible state at time $t$. It depends on all the neighborhood $j\in \mathbb{N}(i)$ that are either in the susceptible state or in the infected state but fail the transmission:
\begin{equation}
	q_{i, t}=\prod_{j \in \mathcal{N}(i)}\left(1-p_{j, t}+\left(1-\beta_{i, j}\right) p_{j, t}\right).
\end{equation}
The parameter $\beta_{i, j} = \beta t_{i,j}$ represents the transmission rate between node $i$ and $j$.

Similarly, Alberto et al. \cite{aleta2020modelling} construct a multi-layer network with a household layer, a workplace and community layer, and a school layer to investigate how testing, contact tracing, and household quarantine affect the disease transmission. Abueg et al. \cite{abueg2021modeling} simulate the contact network in Washington state by three types of sub-networks: fully connected networks for households, small world networks \cite{watts1998collective} for workplaces, schools or social circles, and random networks for random interactions. Nishi et al. \cite{nishi2020network} further investigate how interventions on the structure of contact networks affect the disease transmission. Murphy et al. \cite{murphy2021deep} propose a novel graph convolution network (GCN) model to capture the contagion dynamics in complex contact networks, which is the recent advantage that combines classic epidemiological model with deep learning technology.

From the above discussion, macroscopic models \cite{jia2020population,kraemer2020effect,lai2020effect} only require the empirical epidemic records at the aggregated level, which can be easily obtained in most parts of the world. However, it also forbids them to explain the underlying mechanism due to the over-aggregation of the model parameters, such as the superspreading phenomenon \cite{Lloyd-Smith2005-ao,adam2020clustering}. Microscopic models greatly improve the expressive power of macroscopic models by introducing individual mobility \cite{murphy2021deep,qian2021scaling,aleta2020modelling}, where the heterogeneous contact patterns explain the emergence of such phenomena. However, it also poses a huge demand for mobility data \cite{anwari2021exploring}, which are often difficult to collect in low- and middle-income countries and raises significant privacy concerns~\cite{ahmad2020state}. This dilemma impels the emergence of multi-scale models that achieve a great balance between the expressive power and data demand of the contagion system.

\textbf{Metapopulation model.} A classic solution is the metapopulation model, where the whole simulation region is divided into several sub-regions, where each sub-region maintains a compartmental model (such as SEIR), where an aggregated mobility network connects these sub-regions. The famous global epidemic and mobility (GLEaM) model \cite{balcan2009multiscale,balcan2010modeling} considers the multi-scale mobility of short-scale commuting flows and long-range airline traffic on an SEIR model, which is still inspiring recent studies \cite{zhang2020changes,chinazzi2020effect}. Wu et al. \cite{wu2020nowcasting} propose a metapopulation model using aggregated human mobility data across more than 300 prefecture-level cities in mainland China to investigate how domestic and international movement affect the outbreak of the disease,
where the number of international outbound and inbound air passengers and the daily number of all domestic outbound and inbound travelers are all considered.
Then, according to the mobility network, the sub-regions are connected. The regional mobility data is easier to obtain, which greatly reduces the difficulty of applying the model in real-world scenarios.
Zhou et al. \cite{zhou2020effects} use anonymous mobile phone data to construct an individual-level SEIR model in a metapopulation manner, where the city is divided into multiple regions and runs different SEIR models with different regions, where the metapopulation matrix is incorporated.

Furthermore, Chang et al. \cite{chang2021mobility} propose a bipartite network with time-varying edges between census block groups (CBGs) and points of interest (POIs), which represents the residential areas and venues where the transmission may occur. Based on this highly flexible framework, the authors successfully reproduce the superspreading phenomenon and evaluate the disproportionate impact of the epidemic towards disadvantaged racial and socioeconomic groups.

\textbf{High-order network model.}
Another problem that lies in the single-scale simulation methods is the ignorance of high-order effects during the disease transmission. For example, the macroscopic compartmental models assume a heterogeneous first-order contact between different compartments, and the microscopic models only consider direct connections between individuals. However, such assumptions disobey real-world observations: the contact between family members is closer than the contact between acquaintances, which leads to a higher probability of infection. To tackle this challenge, researchers propose high-order networks, where simplicial complexes \cite{iacopini2019simplicial} or hypergraphs \cite{st2021universal} are introduced to model the transmission patterns in a small group of people. Lacopini et al. \cite{iacopini2019simplicial} assign different infection rates to different motifs to depict the high-order pattern in a social network, where the transition of infection density becomes discontinued and bistable. Similarly, St-Onge et al. \cite{st2021universal} also observe such discontinuous transitions with a different method, where they introduce hyperedges (an edge that connects multiple nodes) to model the high-order feature.

\begin{figure}[t]
	\centering
    \includegraphics[width=0.6\textwidth]{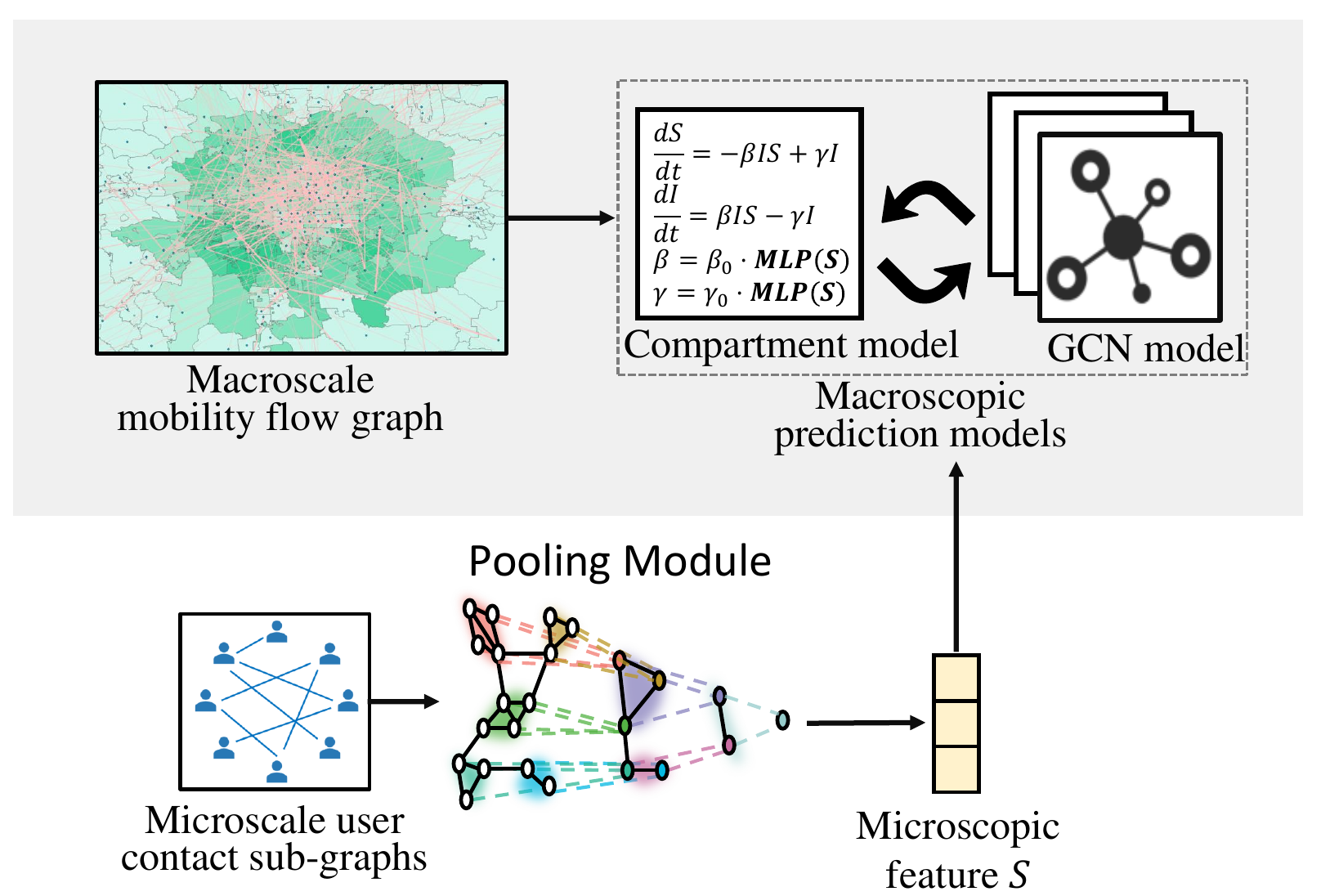}
	\caption{MSDNet~\cite{tang2023enhancing} utilizes multi-scale simulation methods to predict the spatial spread of infectious diseases on the regional scale, where microscopic features derived from microscale user contact graph are incorporated to complete macroscale mechanisms (\emph{Objective B}).}
\label{fig:epidemicExample}
\end{figure}

To better illustrate how to use multi-scale simulation methods and related theories to analyze real-world complex systems, we further present an example in terms of epidemic contagion systems of MSDNet~\cite{tang2023enhancing}. In this example, multi-scale simulation methods are employed to predict the spatial spread of infectious diseases on a regional scale, which is mainly based on the mobility flow between different regions. However, only considering mobility flow between regions, which is regarded as the macroscale information in this scenario, fails to incorporate the intricate details of the mobility behaviors of individual users, such as face-to-face encounters. This limitation consequently results in incomplete macroscopic mechanisms derived by employing machine learning approaches to model the spatial spread of epidemics through mobility flow between different regions. Under such circumstances, it becomes instrumental to incorporate microscale information to complete macroscale mechanisms (\emph{Objective B}). In particular, as shown in Figure~\ref{fig:epidemicExample}, MSDNet extracts the microscopic feature of the user contact sub-graph of each region through graph pooling methods~\cite{ying2018hierarchical}, which is then utilized to modulate key epidemiological parameters in the classical compartment model and also regarded as a ``pre-trained'' feature vector of the region in the macroscale GCN model. The spatial spread of infectious diseases is predicted by collaboratively utilizing the knowledge-driven compartmental model and the data-driven GCN model.

To summarize, going beyond single-scale simulation methods that are stuck in the dilemma of expressive power and data demand, multi-scale simulation uses the metapopulation technique to improve the simulation scale with aggregated level mobility data. It also reveals the underlying high-order effect of disease transmission by considering the contact pattern of a group of people, which brings the simulation of epidemic contagion systems to a new height.

\section{Open Challenges and Future Directions}
A review of the objectives, methods, and applications of multi-scale modeling and simulation for complex systems has been presented in our paper. As we discussed above, unknown mechanisms and expensive computational costs are two key problems in the simulation of complex systems. The former indicates that the mechanism of system evolution is not clear at a specified scale, and the latter refers to the extremely high computational complexity in handling high-dimensional state variables of complex systems. Multi-scale modeling and simulation techniques are regarded as a potential way to solve these problems. To be specific, we can derive the target-scale mechanism by using the knowledge from other scales, which addresses the problem of unclear mechanisms. To deal with the intractable computation, a straightforward solution is to adopt the dimension reduction technique to approximate the massive state variables.
As we surveyed in previous sections, many recent works make efforts to develop different multi-scale methods to address these problems.
However, we still have to face the following challenges:
\begin{packed_itemize}
\item \textbf{How to extract the valuable information at the scale with clear mechanisms for deriving target-scale mechanism}.
The state variables at the scale with clear mechanisms may be extremely high-dimensional, which results in unacceptable computation complexity. At the same time, there exist complex correlations between mechanisms at different scales, which makes it difficult to explore. To further illustrate this challenge, consider the example of traffic optimization in transportation systems. At the microscopic scale, optimizing traffic involves a number of variables, including individual vehicle speed, acceleration, and lane changes, all of which significantly contribute to the overall dynamics of traffic. However, these state variables collectively generate exceptionally high-dimensional data, making computational processing prohibitively expensive. Moreover, there are intricate relationships between traffic mechanisms at different scales. For example, the behavior of individual vehicles on roads can influence macroscopic traffic patterns, such as traffic congestion during rush hours. Thus, it is challenging to extract valuable information to derive the target-scale mechanism.
\item \textbf{How to balance the trade-offs between the computation efficiency and simulation accuracy}.
Ideally, if we were to model each state and their interdependencies within a complex system, we could achieve very high simulation accuracy. However, this requires an enormous amount of computational resources, making it practically unattainable in real-world situations.
Dimension reduction methods are able to reduce the number of variables that should be handled. They can improve the computation efficiency but may lose some simulation accuracy. A simple way mentioned before is to reduce the dimensions in the unimportant areas. However, it is hard to identify which areas are unimportant in complex systems. VAEs~\cite{hernandez2018variational} have emerged as a potential solution to this challenge. VAEs have the capability to reduce complex, nonlinear processes to a low-dimensional embedding with high fidelity. However, it introduces uncertainty when determining the appropriate number of dimensions, and increases the overfitting risks. These factors can influence the simulation results. Thus, how to balance the trade-offs between computation efficiency and simulation accuracy is still challenging.
\item\textbf{How to jointly utilize the transferable information at different scales in system simulation.} In complex systems, the states at different scales interact with each other. Thus, to simulate the states at a specified scale, we not only need to understand the evolution mechanism of this scale, but also need to explore the information from other scales. Meanwhile, the evolution mechanisms at other scales are also affected by the information from this scale. The recently emerged method PINN~\cite{karniadakis2021physics} provides a preliminary attempt toward information transfer across different scales. It integrates known physical laws as constraints during the training process. However, due to its reliance on known physical laws, its capacity for information transfer remains constrained. Thus, it is non-trivial to jointly utilize the transferable information at different scales for simulation.
\end{packed_itemize}
Based on the above challenges, we discuss the following future possible research directions:
\begin{packed_itemize}
\item \textbf{Generalized multi-scale simulation framework}. Most existing works focus on specified methods for multi-scale modeling and simulation in their own research areas. These methods are coupled with the knowledge in the specified areas, which makes it hard to generalize to other areas. A potential way is to construct a generalized multi-scale simulation framework. This framework can establish a new paradigm to guide how useful information is transferred across scales.
\item \textbf{Iterative multi-scale modeling and simulation approaches}. The evolution of state variables at a scale would affect the evolution process at other scales. When simulating a specified scale, we not only consider the influence from other scales, but also explore how this scale acts on other scales. Conducting this iterative interaction allows us to better explore the evolution mechanism, which contributes to the accurate simulation. Thus, developing iterative multi-scale modeling and simulation approaches is a promising direction for future work.
\item \textbf{Data and knowledge joint driven multi-scale simulation}. In multi-scale simulation, data-driven methods rely on the observation data at the corresponding scale. Their effectiveness may be greatly affected in the scenario where the observation data is not covered. Knowledge-based methods have a relatively strong generalization ability, but it is hard to model the evolution mechanism of a scale with massive high-dimensional state variables. Thus, we propose to use the advantages of both methods by integrating data and knowledge in multi-scale simulation. Specifically, it integrates the generalization of knowledge-driven methods and the powerful learning ability of data-driven methods, which benefits exploring the complex evolution mechanism at the corresponding scale. Thus, studying data and knowledge joint driven multi-scale simulation is an important tendency for future research.
\end{packed_itemize}

\section{Conclusion and Summary}
Multi-scale modeling and simulation techniques are playing an increasingly important role in understanding, predicting, and controlling diverse complex systems, which is essential for numerous cutting-edge technology applications. In this survey, we systematically review the literature on multi-scale simulation of major matter systems and social systems, and provide a novel taxonomy in terms of their objectives, methods, and applications. We hope that this interdisciplinary survey will help researchers make better use of existing multi-scale simulation techniques, and also encourage the development of better multi-scale simulation methods.

\bibliographystyle{plain}
\bibliography{sample-base.bib}

\end{document}